\documentclass[%
 reprint,
 amsmath,amssymb,
 aps,
]{revtex4-2}

\usepackage{graphicx}
\usepackage{dcolumn}
\usepackage{bm}

\usepackage{physics}

\usepackage[utf8]{inputenc}
\usepackage{tcolorbox}
\usepackage{amsmath}
\usepackage{amsfonts}
\usepackage{amssymb}
\usepackage{float}
\usepackage{xcolor}
\usepackage{soul}
\usepackage{textcomp}
\usepackage{multirow}
\usepackage{xcolor}
\usepackage{mathtools}
\usepackage{lipsum}
\usepackage{comment}
\usepackage{booktabs}
\usepackage{textcomp}

\usepackage[colorlinks,citecolor=blue,linkcolor=red,anchorcolor=blue,urlcolor=blue]{hyperref}

\begin{document}

\preprint{APS/123-QED}

\title{Fluctuation theorems for autonomous work in the quantum regime}

\author{Xiu-Hua Zhao}
\affiliation{School of Physics, Peking University, Beijing 100871, China}
\author{H. T. Quan}%
\email{htquan@pku.edu.cn}
\affiliation{School of Physics, Peking University, Beijing 100871, China}
\affiliation{Collaborative Innovation Center of Quantum Matter, Beijing 100871, China}
\affiliation{Frontiers Science Center for Nano-Optoelectronics, Peking University, Beijing 100871, China}

\date{\today}

\begin{abstract}
Fluctuation theorems for work provide universal constraints on nonequilibrium fluctuations, yet their quantum generalizations often rely on externally prescribed classical driving protocols. While for classical systems, fluctuation theorems have been extended to autonomous work, where the dynamics of the work source is subject to the backaction of the system, their generalization to the quantum regime is constrained by the uncertainty principle. Here, we extend fluctuation theorems for autonomous work from the classical regime to the quantum regime. By performing successive projective measurements over the work source and the system, we derive Jarzynski-type and Crooks-type fluctuation theorems for autonomous inclusive work from initial mixed thermal states. These relations are analogous to fluctuation theorems for autonomous work in the classical regime and explicitly incorporate the fluctuations of the work source. However, quantum noncommutativity prevents a consistent reduction to the nonautonomous counterparts, even in the limit of a large work source and correspondingly negligible backaction. By contrast, under the exclusive work definition, the nonautonomous limit is recovered when the measured observable of the work source commutes with its bare Hamiltonian and the backaction of the system on the work source is negligible. Our results are illustrated with the Dicke model, where a single-mode radiation field and an ensemble of two-level atoms act as the system of interest and the work source, respectively.
\end{abstract}

\maketitle


\section{Introduction}

With the development of thermodynamics in nonequilibrium stochastic regimes~\cite{sekimotoStochasticEnergetics2010,pelitiStochasticThermodynamicsIntroduction2021,shiraishiIntroductionStochasticThermodynamics2023,seifertStochasticThermodynamics2025}, constraints on nonequilibrium work have evolved from the inequality form of the second law, $\langle w \rangle \ge \Delta F$, to equalities known as fluctuation theorems~\cite{bochkovGeneralTheoryThermal1977,evansProbabilitySecondLaw1993,gallavotti_dynamical_1995_JSP,gallavotti_dynamical_1995_PRL,jarzynskiNonequilibriumEqualityFree1997,crooksEntropyProductionFluctuation1999,liphardtEquilibriumInformationNonequilibrium2002,seifertEntropyProductionStochastic2005,chenHierarchicalStructureFluctuation2023}, such as the Jarzynski equality $\langle e^{-\beta w} \rangle = e^{-\beta \Delta F}$~\cite{jarzynskiNonequilibriumEqualityFree1997}, and the Crooks fluctuation theorem $P^{\text{F}}(w)/P^{\text{B}}(-w) = e^{\beta(w-\Delta F)}$~\cite{crooksEntropyProductionFluctuation1999}, which encodes microscopic reversibility at the level of individual trajectories. Here, $w$ is the stochastic work, $\Delta F$ is the free energy difference, and $\beta=1/(k_{\text{B}}T)$ is the inverse temperature. The superscripts $\text{F}$ and $\text{B}$ denote the forward and the time-reversed processes, respectively. These relations have been established for classical stochastic dynamics and extended to quantum systems with appropriate definitions of work, most commonly based on two-time projective measurements~\cite{kurchan_quantum_2000,tasaki_jarzynski_2000,talknerFluctuationTheoremsWork2007,talkner_tasakicrooks_2007,espositoNonequilibriumFluctuationsFluctuation2009,campisiColloquiumQuantumFluctuation2011,anExperimentalTestQuantum2015,jarzynski_quantum-classical_2015}. Across macroscopic, stochastic and quantum contexts, work is usually defined as the energy change induced by externally controlled parameters, as widely employed in studies of stochastic thermodynamics, quantum thermodynamics, and thermodynamic control~\cite{jarzynskiNonequilibriumEqualityFree1997,sekimotoStochasticEnergetics2010,quanQuantumThermodynamicCycles2007,schmiedlEfficiencyMaximumPower2008,maExperimentalTest12020,liNonequilibriumWorkRelations2023,zhai_quantum_2025}. In this formulation, the system is driven according to a prescribed protocol, and the work parameter, e.g., the coordinate of the work source, is not a dynamical variable. Thermal machines of this type are usually called \textit{nonautonomous}~\cite{quanQuantumThermodynamicCycles2007,campisiColloquiumQuantumFluctuation2011,deffner_information_2013,serra-garcia_mechanical_2016,jarzynskiFluctuationTheoremsAutonomous2025}.

In contrast, for practical thermal machines, the coordinate of the work source is usually a dynamical variable. For example, in a steam engine, the angular coordinate of the flywheel is not externally controlled according to a prescribed protocol but determined by the coupled dynamics of the flywheel and the working substance~\cite{jarzynskiFluctuationTheoremsAutonomous2025,bennett_search_1975}. Due to the coupling between the work source and the system, there is a non-negligible backaction of the system on the work source~\cite{deffner_information_2013,jarzynskiFluctuationTheoremsAutonomous2025}, which is absent in the nonautonomous thermal machines. Nevertheless, when the work source is sufficiently large and rigid, such backaction becomes negligibly small, rendering the work protocol a prescribed protocol and the thermal machine an effective nonautonomous thermal machine~\cite{deffner_information_2013,jarzynskiFluctuationTheoremsAutonomous2025}. When the work source is not sufficiently large, e.g., the system of interest and the work source are comparable, the backaction is non-negligible~\cite{jarzynskiFluctuationTheoremsAutonomous2025,allahverdyan_thomsons_2004,tonner_autonomous_2005,maes_second_2007,deffner_information_2013,frenzel_quasi-autonomous_2016,roulet_autonomous_2017,iino_introduction_2020,jha_jaynes-cummings_2025}. In this case, the system and the work source must be treated as a coupled entity. Within stochastic thermodynamics, Jarzynski and collaborators have successfully extended fluctuation theorems into autonomous settings~\cite{jarzynskiFluctuationTheoremsAutonomous2025}, yielding
\begin{equation}
\label{eq:classical-autonomous-Jarzynski-equality}
    \langle e^{-\beta w + \beta \Delta F - \Delta\phi} \rangle = 1.
\end{equation}
where $\Delta \phi$ denotes the entropy change of the work source along phase-space trajectories. The free energy difference $\Delta F$ becomes a fluctuating quantity that depends on the state of the work source. This relation reduces to the standard fluctuation theorem for nonautonomous work in the limit of a large work source with a sharply localized initial distribution~\cite{jarzynskiFluctuationTheoremsAutonomous2025}. It is desirable to extend the framework of fluctuation theorems for autonomous work from the classical regime to the quantum regime, given the growing interest in quantum autonomous devices~\cite{niedenzu_concepts_2019,elouard_extending_2023,han_quantum_2024,antoniomaringuzmanKeyIssuesReview2024,han_measuring_2025,varmaQuantumMeasurementWork2025,tang_information_2026,cepollaro_autonomization_2026}. However, extending Eq.~\ref{eq:classical-autonomous-Jarzynski-equality} to the quantum regime is nontrivial due to the following difficulties: (i) the notion of trajectory-dependent entropy of the work source fails in the quantum regime; (ii) the quantum uncertainty principle gives rise to intrinsic fluctuations of the work source and imposes fundamental constraints on measurements.

\begin{figure}[!h]
    \centering
    \includegraphics[width=0.98\linewidth]{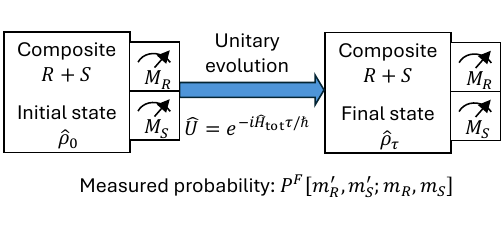}
    \caption{General framework of formulating fluctuation theorems for autonomous work in the quantum regime. The composite system (the work source $R$ and the system of interest $S$) undergoes unitary evolution. Successive projective measurements over the work source and the system are implemented at initial and final times. Here, $m_{R}$ ($m_{R}^{\prime}$) and $m_{S}$ ($m_{S}^{\prime}$) denote the outcomes of the initial (final) measurements over the work source and the system, respectively.}
    \label{fig:theory-framework}
\end{figure}

In this work, we consider a time-independent composite system (a system plus a work source)~\cite{allahverdyan_thomsons_2004,maes_second_2007,deffner_information_2013,jarzynskiFluctuationTheoremsAutonomous2025,tang_information_2026} and formulate fluctuation theorems for autonomous work in the quantum regime based on projective measurements, as shown in Fig.~\ref{fig:theory-framework}. 
We generalize the standard two-point energy measurement scheme~\cite{kurchan_quantum_2000,tasaki_jarzynski_2000,talknerFluctuationTheoremsWork2007,talkner_tasakicrooks_2007} by introducing successive measurements over an observable of the work source and the Hamiltonian of the system of interest, thereby defining inclusive work~\cite{jarzynskiComparisonFarfromequilibriumWork2007} for autonomous systems in the quantum regime. 
Exploiting unitary reversibility and assuming initial thermal equilibrium within each sector of the observable of the work source, we derive fluctuation theorems with a three-term structure analogous to Eq.~\ref{eq:classical-autonomous-Jarzynski-equality}. 
We further show that the wave nature of the work source prevents a consistent reduction of these autonomous relations to their nonautonomous counterparts. 
Alternatively, under the exclusive work definition~\cite{jarzynskiComparisonFarfromequilibriumWork2007}, the fluctuation theorems for autonomous work recover nonautonomous ones when the backaction of the system on the work source is negligible and the measured observable of the work source commutes with its bare Hamiltonian. Our study thus extends fluctuation theorems for autonomous work to fully quantum regimes and establishes an operational framework for investigating quantum autonomous devices.

\section{General Framework}
Considering a system $S$, coupled to a work source $R$, the total time-independent Hamiltonian reads 
\begin{equation}
\label{eq:Hamiltonian-total}
    \hat{H}_{\text{tot}} = \hat{I}_{R} \otimes \hat{H}_{S}^{0} + \hat{A}_{R}\otimes\hat{V}_{S} + \hat{H}_{R}^{0}\otimes\hat{I}_{S},
\end{equation}
where $\hat{H}_{S}^{0}$ and $\hat{H}_{R}^{0}$ denote the bare Hamiltonians of the system of interest and the work source, respectively. $\hat{A}_{R}$ ($\hat{V}_{S}$) denotes an observable of the work source (the system), and $\hat{A}_{R}\otimes\hat{V}_{S}$ describes their interaction. Here, for simplicity, we consider a single factorized coupling $\hat{A}_R \otimes \hat{V}_S$, which is common in quantum physics~\cite{quanDecayLoschmidtEcho2006,schroder_work_2010,aspelmeyerCavityOptomechanics2014,kirtonIntroductionDickeModel2019,felicetti_universal_2020}. $\hat{I}_{R}$ and $\hat{I}_{S}$ are identity operators on the respective local Hilbert spaces. In accordance with the definition of inclusive work, where the interaction term in Eq.~\ref{eq:Hamiltonian-total} is included in the system energy, we are interested in the following Hamiltonian,
\begin{equation}
    \hat{H}_{SR} \equiv \hat{I}_{R} \otimes \hat{H}_{S}^{0} + \hat{A}_{R}\otimes\hat{V}_{S}.
\end{equation}
When $\hat{H}_{SR}$ is not conserved, there is autonomous energy exchange between the system and the work source along the unitary evolution from the initial state $\hat{\rho}_{0}$ of the composite system to the subsequent state $\hat{\rho}_{t}=\hat{U}(t,0)\hat{\rho}_{0}\hat{U}^{\dagger}(t,0)$. Here, $\hat{U}(t,0)$ is the unitary evolution operator of the composite system that satisfies
\begin{equation}
    \frac{d}{dt}\hat{U}(t,0) = - \frac{i}{\hbar} \hat{H}_{\text{tot}}\hat{U}(t,0).
\end{equation}

Quantum fluctuation theorems for work are fundamentally connected to the reversibility of unitary evolution. Therefore, it is important to determine the relevant transitions between states in Hilbert space. In the nonautonomous scenario, projective measurements of the system Hamiltonian determine these transitions by projecting the system onto energy eigenstates, up to possible degeneracy indices~\cite{espositoNonequilibriumFluctuationsFluctuation2009}. In the autonomous scenario considered here, however, one must additionally determine the state of the work source, which governs the interaction experienced by the system. To this end, we perform successive projective measurements over the observable $\hat{A}_{R}$ of the work source and the Hamiltonian $\hat{H}_{SR}$. The corresponding projector is defined as 
\begin{equation}
    \hat{\Pi}^{a}_{n} \equiv \ket{a, E_{n}^{a}}\bra{a, E_{n}^{a}} =  \ket{a}\bra{a} \otimes \ket{E_{n}^{a}}\bra{E_{n}^{a}},
\end{equation}
where $a$ is the eigenvalue of $\hat{A}_{R}$ and $E_{n}^{a}$ is the eigenvalue of the conditional Hamiltonian of the system $\hat{H}_{S}(a)\equiv \hat{H}_{S}^{0} + a\hat{V}_{S}$. This reflects the fact that, upon measuring $\hat{A}_{R}$, the operator $\hat{H}_{SR}$ reduces to an effective Hamiltonian acting solely on the local Hilbert space of the system. Here we assume that $\hat{A}_{R}$ is a discrete-spectrum operator so that projecting the work source to a definite sector denoted by $a$ will not lead to singularities. The joint eigenbasis ${\ket{a, E_{n}^{a}}}$ satisfies orthonormal conditions $\langle {a^{\prime}, E_{m}^{a^{\prime}}} | {a, E_{n}^{a}} \rangle = \delta_{aa^{\prime}}\delta_{mn}$ and $\sum_{a,n}\ket{a, E_{n}^{a}}\bra{a, E_{n}^{a}} = \hat{I}_{R}\otimes\hat{I}_{S}$. Here, for simplicity without loss of generality, we assume that $\ket{a}$ and $\ket{E_{n}^{a}}$ correspond to nondegenerate eigenvalues in their respective local Hilbert spaces. The joint probability of obtaining the measurement outcome $|a, E_{n}^{a}\rangle$ at the initial time $t=0$ and $|a^{\prime}, E_{m}^{a^{\prime}}\rangle$ at the final time $t=\tau$ is given by
\begin{equation}
\label{eq:joint_prob_general_forward_trace}
    P^\text{F}[\Gamma] \equiv  \mathrm{Tr} \left\{ \hat{\Pi}^{a^{\prime}}_{m} \hat{U}(\tau,0) \hat{\Pi}^{a}_{n} \hat{\rho}_{0} \hat{\Pi}^{a}_{n} \hat{U}^{\dagger}(\tau,0)\hat{\Pi}^{a^{\prime}}_{m} \right\},
\end{equation}
Here, $\Gamma$ denotes the transition from $|a, E_{n}^{a}\rangle$ to $|a^{\prime}, E_{m}^{a^{\prime}}\rangle$. $P^{\text{F}}[\Gamma]$ satisfies the normalization condition $\sum_{\Gamma}P^{\text{F}}[\Gamma]=\sum_{a^{\prime}, m; a, n}P^{\text{F}}[\Gamma]=1$. $\hat{\rho}_{0}$ is the initial mixed thermal state~\cite{initial_state}, which commutes with $\hat{\Pi}_{n}^{a}$,
\begin{equation}
    \label{eq:initial_state_general_forward}
    \hat{\rho}_{0} = \sum_{a}p_a \left[\ket{a}\bra{a}\otimes e^{-\beta \hat{H}_{S}(a)}Z_{S}^{-1}(a) \right],
\end{equation}
with $Z_{S}(a)\equiv \mathrm{Tr}_{S}\{\mathrm{exp} [-\beta \hat{H}_{S}(a)]\}$. 
$p_{a}$ is the probability of sampling $a$, which can be arbitrary as long as it is normalized~\cite{jarzynskiFluctuationTheoremsAutonomous2025}. The inclusive work performed by the work source on the system is defined as the measured energy difference,
\begin{equation}
\label{eq:work_definition}
    w[\Gamma] \equiv E_{m}^{a^{\prime}} - E_{n}^{a},
\end{equation}
which depends on the state of the work source.

Then let us consider the time-reversed process, where the reversal is performed at $t=\tau$. The evolution operator becomes $\hat{U}_{\text{B}}(t,0) = \hat{\Theta} \hat{U}(\tau-t, \tau)\hat{\Theta}^{\dagger}$, where $\hat{\Theta}$ is the time-reversal operator satisfying $\hat{\Theta}\hat{\Theta}^{\dagger}=1$. The initial state of the backward process is denoted as $\hat{\rho}_{0}^{\text{B}}$. 
Defining $\hat{\rho}_{0}^{\text{tr}} \equiv \hat{\Theta}^{\dagger}\hat{\rho}_{0}^{\text{B}}\hat{\Theta}$, we obtain the following joint probability for the backward process (see details in Appendix~\ref{appendix:time-reversal}) 
\begin{equation}
\label{eq:joint_prob_general_backward_trace}
    P^{\text{B}}[\tilde{\Gamma}] = \mathrm{Tr} \left\{ \hat{\Pi}^{a}_{n} \hat{U}^{\dagger}(\tau,0) \hat{\Pi}^{a^{\prime}}_{m} \hat{\rho}^{\text{tr}}_{0} \hat{\Pi}^{a^{\prime}}_{m} \hat{U}(\tau,0)\hat{\Pi}^{a}_{n} \right\}.
\end{equation}
$\tilde{\Gamma}$ denotes the transition from $\hat{\Theta}|a^{\prime}, E_{m}^{a^{\prime}}\rangle$ to $\hat{\Theta}|a, E_{n}^{a}\rangle$. $\hat{\Theta}\ket{a, E_{n}^{a}}$ is the simultaneous eigenstate of $\hat{\Theta}\hat{A}_{R}\hat{\Theta}^{\dagger}$ with the eigenvalue $a$ and $\hat{\Theta}\hat{H}_{S}(a)\hat{\Theta}^{\dagger}$ with the eigenvalue $E_n^{a}$. Here the eigenvalues remain the same as the forward process because they are real numbers. $P^{\text{B}}[\tilde{\Gamma}]$ is normalized as $\sum_{\tilde{\Gamma}}P^{B}[\tilde{\Gamma}]=\sum_{a, n; a^{\prime}, m}P^{B}[\tilde{\Gamma}]=1$. The time-reversed initial state of the backward process reads
\begin{equation}
\label{eq:initial_state_general_backward}
    \hat{\rho}_{0}^{\text{tr}} = \sum_{a^{\prime}} p^{\text{tr}}_{a^{\prime}} \left[\ket{a^{\prime}}\bra{a^{\prime}}\otimes e^{-\beta \hat{H}_{S}(a^{\prime})}Z_{S}^{-1}(a^{\prime}) \right],
\end{equation}
where $p_{a^{\prime}}^{\text{tr}}$ denotes the initial probability of sampling $a^{\prime}$ for the observable $\hat{A}_{R}$ of the work source in the backward process. 

To formulate fluctuation theorems for autonomous work, we are interested in the probability distribution of the following entropy-like quantity~\cite{maesOriginUseFluctuation2003,seifertEntropyProductionStochastic2005,chenHierarchicalStructureFluctuation2023},
\begin{equation}
\label{eq:ratio-prob}
\begin{aligned}
    R[\Gamma] \equiv \ln \frac{P^{\text{F}}[\Gamma]}{P^{\text{B}}[\tilde{\Gamma}]},
\end{aligned}
\end{equation}
which is related to the transitions from $\ket{a, E^{a}_{n}}$ to $|a^{\prime}, E^{a^{\prime}}_{m}\rangle$ in the forward process and from $\hat{\Theta}|a^{\prime}, E^{a^{\prime}}_{m}\rangle$ to $\hat{\Theta}\ket{a, E^{a}_{n}}$ in the backward process. The detailed meaning of the above quantity can be further clarified by substituting the expressions of $P^{\text{F}}[\Gamma]$ and $P^{\text{B}}[\tilde{\Gamma}]$, as well as the initial states Eqs.~\ref{eq:initial_state_general_forward} and \ref{eq:initial_state_general_backward} into Eq.~\ref{eq:ratio-prob}, yielding,
\begin{equation}
\label{eq:total_random_variable_general}
R[\Gamma]= \beta w[\Gamma] -  \beta \Delta F_{S}[\Gamma] + \Delta\phi[\Gamma],
\end{equation}
where
\begin{align}
    & \Delta F_{S}[\Gamma] \equiv F_{S}(a^{\prime}) - F_{S}(a),\\
    &\Delta\phi[\Gamma] \equiv \phi_{a^{\prime}}^{\text{tr}} - \phi_{a},
\end{align}
with $F_{S}(a) \equiv -\beta^{-1}\ln Z_{S}(a)$, $\phi_{a} \equiv - \ln p_{a}$ and $\phi_{a^{\prime}}^{\text{tr}} \equiv - \ln p_{a^{\prime}}^{\text{tr}}$. The initial probability $p_{a^{\prime}}^{\text{tr}}$ in the backward process can, in principle, be chosen arbitrarily. Here, it is chosen to coincide with the probability of obtaining $a^{\prime}$ from the measurement of the observable $\hat{A}_{R}$ at the end of the forward process, so that $\Delta\phi[\Gamma]$ represents the change in the measurement statistics of the work source during the forward process. Consequently, the three terms on the right-hand side of Eq.~\ref{eq:total_random_variable_general} have clear physical meanings: the first corresponds to the work performed on the system during the evolution from $\ket{a, E^{a}_{n}}$ to $|a^{\prime}, E^{a^{\prime}}_{m}\rangle$, the second corresponds to the free energy change, and the third characterizes the distribution change of the work source. The probability distribution of the quantity $R[\Gamma]$ in the forward process is calculated as
\begin{equation}
    P^{\text{F}}(R) \equiv \sum_{a^{\prime}, m; a, n} P^{\text{F}}[\Gamma]\;\delta(R[\Gamma]-R).
\end{equation}
which gives the Jarzynski-type fluctuation theorem as follows,
\begin{equation}
\label{eq:Jarzynski_general}
    \langle e^{-R} \rangle_{\text{F}} = 1,
\end{equation}
where $\langle\cdot\rangle_{\text{F}}$ denotes an average over all realizations of the forward process. In the above equality, $R$ is the value of $R[\Gamma]$ (Eq.~\ref{eq:total_random_variable_general}), which shares the same structure as the exponent in Eq.~\ref{eq:classical-autonomous-Jarzynski-equality}. Compared to the result Eq.~\ref{eq:classical-autonomous-Jarzynski-equality} in the classical regime, the fluctuation information of the work source is encoded here in the initially and finally measured probabilities in the representation of the observable $\hat{A}_{R}$, rather than in continuous phase-space trajectories. In the quantum regime, successive measurements of the work source and the system play an important role in deriving the fluctuation theorem for autonomous work. They determine the microscopic states of the composite system and enable a consistent definition of the system’s free energy.

The Crooks-type fluctuation theorem for autonomous work is obtained by combining the probability distributions of the entropy-like quantity $R$ in the forward and backward processes. For the backward evolution, $R$ is the value of $R[\tilde{\Gamma}] \equiv \ln P^{\text{B}}[\tilde{\Gamma}] / P^{\text{F}}[\Gamma]$, which is equal to $-R[\Gamma]$, and its probability distribution in the backward process reads 
\begin{equation}
    P^{\text{B}}(R) \equiv \sum_{a, n; a^{\prime}, m} P^{\text{B}}[\tilde{\Gamma}]\;\delta(R[\tilde{\Gamma}]-R).
\end{equation}
With the expressions of $P^{\text{F}}(R)$ and $P^{\text{B}}(R)$, it can be verified that 
\begin{equation}
\label{eq:Crooks_general}
    \frac{P^{\text{F}}(R)}{P^{\text{B}}(- R)} = e^{R} .
\end{equation}

The fluctuation theorems can be equivalently expressed with generating functions~\cite{talkner_tasakicrooks_2007,espositoNonequilibriumFluctuationsFluctuation2009,campisiColloquiumQuantumFluctuation2011}. Define generating functions in the forward and backward processes respectively as
\begin{equation}
\label{eq:generating-function-forward}
    G^{\text{F}}_{R}(u) \equiv \sum_{R}e^{iuR}P^{\text{F}}(R) = \mathrm{Tr} \left\{  (\hat{\rho}_{0}^{\text{tr}})^{-iu} (\hat{\rho}_{\tau})^{1+iu} \right\},
\end{equation}
and
\begin{equation}
    G^{\text{B}}_{R}(u) \equiv \sum_{R}e^{iuR}P^{\text{B}}(R) = \mathrm{Tr} \left\{  (\hat{\rho}_{0})^{-iu}  (\hat{\rho}_{\tau}^{\text{tr}})^{1+iu}  \right\},
\end{equation}
where $\hat{\rho}_{\tau}\equiv \hat{U}(\tau,0)\hat{\rho}_{0}\hat{U}^{\dagger}(\tau,0)$ and $\hat{\rho}_{\tau}^{\text{tr}}\equiv \hat{U}^{\dagger}(\tau,0)\hat{\rho}_{0}^{\text{tr}}\hat{U}(\tau,0)$. We have used the condition that $\hat{\rho}_{0}$ and $\hat{\rho}_{0}^{\text{tr}}$ commute with the projectors. The relation between $G_{R}^{\text{F}}(u)$ and $G_{R}^{\text{B}}(u)$ reads,
\begin{equation}
G^{\text{B}}_{R}(i-u) = G^{\text{F}}_{R}(u).
\end{equation}
Its consistency with Eqs.~\ref{eq:Jarzynski_general} and \ref{eq:Crooks_general} can be verified by setting $u=i$ and by taking the inverse Fourier transform of the generating functions, respectively. 


Defining the cumulant generating function $K_{R}^{\text{F}}(u) \equiv \ln G_{R}^{\text{F}}(u)$ and expanding it around $u=0$ as $K_{R}^{\text{F}}(u) = iu \langle R \rangle_{\text{F}} - \sigma_{R}^{\text{F} \; 2}u^2/2 + o(u^2)$, where $\sigma_{R}^{\text{F}}$ is the standard deviation of $R$ in the forward process, we obtain
\begin{equation}
\label{eq:fluctuation-dissipation-relation}
    \beta (\langle w - \Delta F_{S}\rangle_{\text{F}}) + \langle\Delta\phi\rangle_{\text{F}} \approx \frac{1}{2}\sigma_{R}^{\text{F} \;2}.
\end{equation}
We have used $K_{R}^{\text{F}}(i)=\ln G_{R}^{\text{F}}(i)=0$. Equation~\ref{eq:fluctuation-dissipation-relation} is the fluctuation-dissipation relation for the autonomous work in the near-equilibrium regime where higher-order fluctuations of $R$ are neglected. 
This relation will formally reduce to its nonautonomous counterpart $\langle w \rangle_{\text{F}} - \Delta F_{S} = \beta \sigma_{w}^{\text{F}\; 2}/2$~\cite{jarzynskiNonequilibriumEqualityFree1997}, where $\sigma_{w}^{\text{F}\; 2} = \langle w^2 \rangle_{\text{F}} - \langle w \rangle_{\text{F}}^{2}$, if the probability change $\Delta \phi$ of the work source vanishes and the free energy change is a definite number rather than a fluctuating quantity. However, quantum noncommutativity hinders the existence of such a nonautonomous limit, which will be discussed in the section \textit{Quantum Obstruction to the Nonautonomous Limit}.

\section{Demonstration With the Dicke Model}

\begin{figure}[!h]
    \centering
    \includegraphics{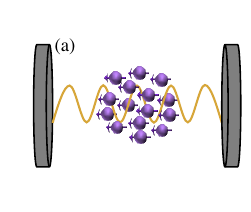}
    \includegraphics{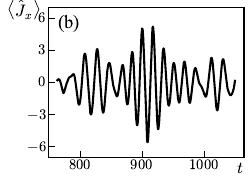}
    \includegraphics{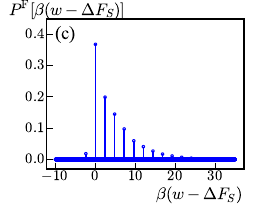}
    \includegraphics{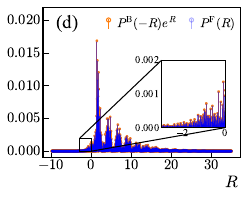}
    \caption{(a) Illustration of the Dicke model, consisting of a single-mode radiation field and $N$ two-level atoms. (b) The evolution of the expected value $\langle\hat{J}_{x}\rangle$ of the atoms. (c) and (d) show the statistics of the finite-time excitation $\beta(w-\Delta F_{S})$ of the radiation field and the joint fluctuation variable $R=\beta(w-\Delta F_{S})+\Delta\phi$, respectively. In panels (b) (c) and (d), $\hbar=1,\beta=3,\omega_c = 0.8, \omega_z = 0.4, \lambda = 0.25$. The number of atoms is $N = 50$. The dimension of the Hilbert space of the radiation field is truncated at $50$, a large enough value in the evolution. The initial distribution $p_{j}$ of atoms is generated by the normal distribution with the average value $-25$ and the standard deviation $5$ and then renormalized. The initial state of the radiation field is the mixed thermal state for each $j$ (Eq.~\ref{eq:initial_state_general_forward}). The time to obtain panels (c) and (d) is $\tau=1000$.}
    \label{fig:Dicke-general}
\end{figure}

We employ the Dicke model~\cite{kirtonIntroductionDickeModel2019}, as illustrated in Fig.~\ref{fig:Dicke-general}(a), to demonstrate the fluctuation theorems for the autonomous work. The Hamiltonian of the Dicke model, consisting of a single-mode radiation field and $N$ two-level atoms, reads
\begin{equation}
\label{eq:Hamiltonian-Dicke-model}
    \hat{H}_{\text{tot}} = \hbar\omega_{c}\hat{I}_{\text{am}} \otimes \hat{a}^{\dagger}\hat{a} + \frac{2\lambda}{\sqrt{N}}\hat{J}_{x} \otimes \left(\hat{a}^{\dagger} + \hat{a}\right) + \omega_{z} \hat{J}_{z} \otimes \hat{I}_{\text{rf}},
\end{equation}
where $\hat{J}_{x} \equiv (\hbar/2)\sum_{i=1}^{N}\hat{\sigma}_{i,x}$, $\hat{J}_{z} \equiv (\hbar/2)\sum_{i=1}^{N}\hat{\sigma}_{i,z}$, and $\hat{\sigma}_{i,x(z)}$ denote the Pauli operators. The parameters $\omega_{c}$ and $\omega_{z}$ represent the intrinsic frequencies of the radiation field and the atoms, respectively, while $\lambda$ characterizes the interaction strength. $\hat{I}_{\text{am}}$ ($\hat{I}_{\text{rf}}$) is the identity operator on the local Hilbert space of the atoms (radiation field). $\hat{a}^{\dagger}$ and $\hat{a}$ are creation and annihilation operators of the radiation field. We treat the radiation field as the system of interest and the ensemble of two-level atoms as the work source. The Hamiltonian we are interested in and the conditional Hamiltonian of the system corresponding to a given eigenvalue $j$ of $\hat{J}_{x}$ are respectively given by
\begin{align}
    & \hat{H}_{SR} = \hbar\omega_{c} \hat{I}_{\text{am}}\otimes\hat{a}^{\dagger}\hat{a} + \frac{2\lambda}{\sqrt{N}}\hat{J}_{x} \otimes \left(\hat{a}^{\dagger} + \hat{a}\right),\\
    & \hat{H}_{S}(j) = \hbar\omega_{c} \hat{a}^{\dagger}\hat{a} + \frac{2\lambda}{\sqrt{N}} j \left(\hat{a}^{\dagger} + \hat{a}\right).
\end{align}
Here $\hat{J}_{x}$ serves as the operator $\hat{A}_{R}$ of the work source in the general framework. Owing to the conservation of the total squared angular momentum $\hat{J}^{2} \equiv \hat{J}_{x}^{2} + \hat{J}_{y}^{2} + \hat{J}_{z}^{2}$, we restrict our analysis, for simplicity, to the subspace with the eigenvalue of $\hat{J}^{2}$ being $(N/2)(N/2+1)\hbar^2$. Within this $(N+1)$-dimensional subspace, the eigenvalues of $\hat{J}_{x}$ are $j =-\frac{N}{2}\hbar$, $-\frac{N-2}{2}\hbar$, $-\frac{N-4}{2}\hbar$, \dots, $\frac{N-4}{2}\hbar$, $\frac{N-2}{2}\hbar$, $\frac{N}{2}\hbar$. 

In this model, the initial state reads $\hat{\rho}_{0} = \sum_{j}p_{j} [\ket{j}\bra{j}\otimes e^{-\beta \hat{H}_{S}(j)} Z_{S}^{-1}(j) ]$, with the partition function $Z_{S}(j) = \exp[4\beta\lambda^2j^2/(\hbar\omega_{c}N)]/[1-\exp(-\beta\hbar\omega_{c})]$. The eigenvalues of $\hat{H}_{S}(j)$ are given by $E^{j}_{n} = n \hbar\omega_{c} - 4\lambda^2j^2/(\hbar\omega_{c}N)$, where $n=0,1,2,\dots$ Due to the equally spaced energy spectrum, the difference between the work and the free energy change during the transition $\Gamma$ (from $| j,E_{n}^{j} \rangle$ to $| j^{\prime}, E_{m}^{j^{\prime}} \rangle$) reduces to $w[\Gamma] - \Delta F_{S}[\Gamma] = (m-n)\hbar\omega_{c}$, which corresponds to the transition between different energy levels of the radiation field. By definition, the entropy-like quantities entering the fluctuation theorems, Eqs.~\ref{eq:Jarzynski_general} and \ref{eq:Crooks_general}, are expressed as $R[\Gamma] = (m-n)\beta\hbar\omega_{c} + \Delta\phi[\Gamma]$ for the forward transition $\Gamma$ and $R[\tilde{\Gamma}]=-R[\Gamma]$ for the corresponding backward transition $\tilde{\Gamma}$, where $\Delta\phi[\Gamma] = -\ln p_{j^{\prime}}^{\text{tr}} + \ln p_{j}$. Here, $p_{j^{\prime}}^{\text{tr}}$ is chosen to be the finally measured distribution of $\hat{J}_{x}$ in the forward process and $\Delta\phi[\Gamma]$ represents the distribution change of $\hat{J}_{x}$ in the forward process.

Panels (b-d) in Fig.~\ref{fig:Dicke-general} illustrate the dynamical evolution and statistical properties under the Hamiltonian Eq.~\ref{eq:Hamiltonian-Dicke-model}. In the numerical calculation (see details in Appendix~\ref{appendix:numerical-method-Dicke-inclusive}), we choose a representation in which the total Hamiltonian $\hat{H}_{\text{tot}}$ is real and the time-reversal operator $\hat{\Theta}$ is implemented as complex conjugation~\cite{bakemeierDynamicsDickeModel2013}. Figure~\ref{fig:Dicke-general}(b) illustrates the backaction of the radiation field on the atomic dynamics. If there were no backaction, $\langle \hat{J}_{x}\rangle$ would undergo simple oscillations with frequency $\omega_{z}$. The modulation of both amplitude and frequency observed in Fig.~\ref{fig:Dicke-general}(b) clearly indicates the influence of the radiation field. Figures~\ref{fig:Dicke-general}(c) and (d) show, respectively, the probability distributions of $\beta(w-\Delta F_{S})$ and $R=\beta(w-\Delta F_{S})+\Delta\phi$ at a representative time. The quantity $\beta(w-\Delta F_{S})$ takes discrete values, and its nonzero values with finite probability correspond to transitions between different energy levels of the radiation field induced by the atomic driving. Upon incorporating the distribution change of the atoms, $\Delta\phi$, these transitions satisfy the Crooks-type fluctuation theorems. Specifically, the two distributions in Fig.~\ref{fig:Dicke-general}(d), $P^{\text{F}}(R)$ and $P^{\text{B}}(-R)e^{R}$, collapse onto each other, confirming the validity of the Crooks-type fluctuation theorem Eq.~\ref{eq:Crooks_general}, which in turn implies the corresponding Jarzynski-type relation Eq.~\ref{eq:Jarzynski_general}. 


\section{Quantum Obstruction to the Nonautonomous Limit}
\begin{figure}[!h]
    \centering
    \includegraphics{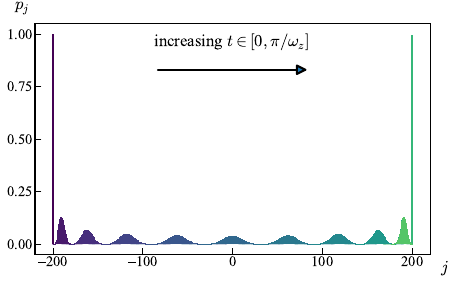}
    \caption{The distribution of $\hat{J}_{x}$ in the Dicke model (Eq.~\ref{eq:Hamiltonian-Dicke-model}) at different times during the evolution in the limit of vanishing backaction. The eleven distributions from the left to the right correspond to $t\in[0,1]$ with the interval $0.1$. Parameters: $\hbar=1,\beta=1,\omega_c = 8, \omega_z = \pi, \lambda = 0.6$. The number of atoms is $N = 400$. The dimension of the Hilbert space of the radiation field is truncated at $20$, a large enough value in the evolution. The initial state of the ensemble of atoms is the eigenstate of $\hat{J}_{x}$ with the minimum eigenvalue, $j=-200$. The initial state of the radiation field is the thermal state with $j=-200$ (Eq.~\ref{eq:initial_state_general_forward}). }
    \label{fig:Dicke-semidecoupling-evolution-PJx}
\end{figure}

In the classical regime, the fluctuation theorems for autonomous work reduce to their nonautonomous counterparts under two conditions~\cite{deffner_information_2013,jarzynskiFluctuationTheoremsAutonomous2025}: 
(i) the work source is in the large-mass limit, such that the backaction of the system on the work source is negligible. In this case, the Hamiltonian evolution of the work source preserves its entropy along trajectories; (ii) the variable of the work source that couples the system and the work source is no longer stochastic, but instead takes a definite value at any time. In the quantum regime, however, such a limiting regime does not exist. Energy exchange between the system and the work source requires $[\hat{H}_{\text{tot}}, \hat{H}_{SR}] \neq 0$. Since $[\hat{H}_{\text{tot}}, \hat{H}_{SR}] = [\hat{H}_{R}^{0}, \hat{A}_{R}]\otimes\hat{V}_{S}$ (using Eq.~\ref{eq:Hamiltonian-total}), this condition implies that the observable $\hat{A}_{R}$ does not commute with the bare Hamiltonian of the work source $\hat{H}_{R}^{0}$. 
This noncommutativity generally leads to nonvanishing $\Delta\phi[\Gamma]$ in Eq.~\ref{eq:total_random_variable_general}. 
Figure~\ref{fig:Dicke-semidecoupling-evolution-PJx} provides explicit evidence that an initial sharp distribution of the observable of the work source inevitably broadens even in the limit of vanishing backaction (see details in Appendix~\ref{appendix:vanishing-backaction-Dicke}). Therefore, in the quantum regime, the observable $\hat{A}_{R}$ cannot evolve deterministically as a classical control parameter. The fluctuation theorems for nonautonomous work are recovered only under a further semiclassical factorization, where the expectation value of $\hat{A}_{R}$ acts as a control parameter in the effective Hamiltonian $\hat{H}_{S}^{\text{sc}}(t) \equiv \hat{H}_{S}^{0} + \langle \hat{A}_{R} \rangle _{t} \hat{V}_{S}$ of the system (see Appendix~\ref{appendix:factorized-semiclassical}).

\section{Fluctuation Theorems for the Exclusive Work}

Replacing inclusive work with exclusive work~\cite{jarzynskiComparisonFarfromequilibriumWork2007}, where the interaction term is attributed to the work source rather than the system of interest, we can likewise establish the detailed and Bochkov–Kuzovlev-type fluctuation theorems~\cite{bochkovGeneralTheoryThermal1977,campisi_quantum_2011,campisiColloquiumQuantumFluctuation2011, jarzynskiFluctuationTheoremsAutonomous2025}. In this exclusive case, since the interaction term is excluded from the system energy, to determine the state of the composite system, the measured observable of the work source is no longer tied to the interaction term. Consequently, in the limit of vanishing backaction, the distribution change of the work source vanishes if the measured observable commutes with the bare Hamiltonian of the work source. This allows the nonautonomous limit~\cite{bochkovGeneralTheoryThermal1977,campisi_quantum_2011,campisiColloquiumQuantumFluctuation2011, jarzynskiFluctuationTheoremsAutonomous2025} to be consistently recovered in the fluctuation theorems for exclusive work.

\subsection{General Framework}
With the definition of exclusive work, in the total Hamiltonian $\hat{H}_{\text{tot}} = \hat{I}_{R}\otimes\hat{H}_{S}^{0} + \hat{A}_{R}\otimes\hat{V}_{S} + \hat{H}_{R}^{0}\times\hat{I}_{S}$, the system energy is $\hat{I}_{R}\otimes \hat{H}_{S}^{0}$, while the other terms $\hat{A}_{R}\otimes\hat{V}_{S} + \hat{H}_{R}^{0}\otimes\hat{I}_{S}$ are attributed to the work source. The eigenstate of $\hat{H}_{S}^{0}$ is $\ket{e_{n}}$ with the eigenvalue $e_{n}$. We initially and finally perform joint measurements over $\hat{\Xi}_{R}$ and $\hat{H}_{S}^{0}$. Here, $\hat{\Xi}_{R}$ is an arbitrary observable on the local Hilbert space of the work source. The joint projector reads
\begin{equation}
    \hat{\pi}_{n}^{\xi} \equiv \ket{\xi,e_{n}}\bra{\xi,e_{n}} = \ket{\xi}\bra{\xi}\otimes\ket{e_{n}}\bra{e_{n}},
\end{equation}
where $\ket{\xi}$ is the eigenstate of $\hat{\Xi}_{R}$ with the eigenvalue $\xi$. Consider the following initial state,
\begin{equation}
    \hat{\eta}_{0}  = \left[\sum_{\xi} p_{\xi} \ket{\xi}\bra{\xi}\right] \otimes  \left[e^{-\beta \hat{H}_{S}^{0}}Z_{0}^{-1}\right],
\end{equation}
where $p_{\xi}$ is the probability of sampling $\xi$ and $Z_{0} \equiv \mathrm{Tr}_{S}[ \text{exp} (-\beta \hat{H}_{S}^{0})]$. For the forward evolution, the joint measurement probability corresponding to the transition $\gamma$ from $\ket{\xi,e_{n}}$ to $\ket{\xi^{\prime},e_{m}}$ is
\begin{equation}
    P^{\text{F}}[\gamma] 
    = \left| \bra{\xi^{\prime},e_{m}}\hat{U}(\tau,0)\ket{\xi,e_{n}} \right|^{2} \bra{\xi, e_{n}}\hat{\eta}_{0}\ket{\xi, e_{n}}.
\end{equation}
We have used the condition that $\langle\xi^{\prime}, e_{m}|\xi,e_{n}\rangle = \delta_{\xi^{\prime}\xi}\delta_{mn}$ and $\sum_{\xi,n}\ket{\xi,e_{n}}\bra{\xi,e_{n}}=\hat{I}_{R}\otimes\hat{I}_{S}$. For the time-reversed process, we perform similar joint measurements at the initial and final times with the time-reversed projector $\hat{\Theta}\hat{\pi}_{n}^{\xi}\hat{\Theta}^{\dagger}$. The joint probability corresponding to the transition $\tilde{\gamma}$ from $\hat{\Theta} |\xi^{\prime}, e_{m}\rangle$ to $\hat{\Theta}|\xi, e_{n}\rangle$ in the backward process is
\begin{equation}
    P^{\text{B}}[\tilde{\gamma}] 
    = \left| \bra{\xi^{\prime},e_{m}}\hat{U}(\tau,0)\ket{\xi,e_{n}} \right|^{2} \bra{\xi^{\prime}, e_{m}}\hat{\eta}_{0}^{\text{tr}}\ket{\xi^{\prime}, e_{m}},
\end{equation}
where 
\begin{equation}
    \hat{\eta}_{0}^{\text{tr}} \equiv \hat{\Theta}^{\dagger}\hat{\eta}_{0}^{\text{B}} \hat{\Theta}  = \sum_{\xi^{\prime}} p_{\xi^{\prime}}^{\text{tr}} \ket{\xi^{\prime}}\bra{\xi^{\prime}}\;\otimes\; e^{-\beta \hat{H}_{S}^{0}}Z_{0}^{-1},
\end{equation}
and $\hat{\eta}_{0}^{\text{B}}$ is the initial state of the backward process. Defining $r[\gamma] \equiv \ln (P^{\text{F}}[\gamma]/P^{\text{B}}[\tilde{\gamma}])$ and $r[\tilde{\gamma}] \equiv \ln (P^{\text{B}}[\tilde{\gamma}]/P^{\text{F}}[\gamma]) = - r[\gamma]$, we find that 
\begin{equation}
    \frac{P^{\text{F}}(r)}{P^{\text{B}}(-r)} = e^{r} , \quad \langle e^{-r} \rangle _{\text{F}} = 1,
\end{equation}
where $P^{\text{F}}(r) = \sum_{\xi^{\prime},m;\xi,n}P^{\text{F}}[\gamma]\delta\left(r[\gamma] - r\right)$, $P^{\text{B}}(r) = \sum_{\xi,n; \xi^{\prime},m}P^{\text{B}}[\tilde{\gamma}]\delta\left(r[\tilde{\gamma}] - r\right)$, and $\langle e^{-r} \rangle_{\text{F}} = \sum_{r}e^{-r}P^{\text{F}}(r)$. Furthermore, the detailed expression of $r[\gamma]$ reads
\begin{equation}
\label{SMeq:random-variable-rF}
    r[\gamma] = \beta w_{\text{exc}}[\gamma] - \ln\frac{p_{\xi^{\prime}}^{\text{tr}}}{p_{\xi}},
\end{equation}
where
\begin{equation}
    w_{\text{exc}}[\gamma] \equiv e_{m} - e_{n}
\end{equation}
is the exclusive work performed on the system during the transition from $|\xi,e_{n}\rangle$ to $|\xi^{\prime},e_{m}\rangle$. The probability $p_{\xi^{\prime}}^{\text{tr}}$ is in principle arbitrary and can be chosen as the measured distribution of $\hat{\Xi}_{R}$ on the final state in the forward process. Then $- \text{ln} (p_{\xi^{\prime}}^{\text{tr}} / p_{\xi})$ represents the entropy change of the work source related to the distribution on $\hat{\Xi}_{R}$. A demonstration of the fluctuation theorems for autonomous exclusive work using the Dicke model can be found in Appendix~\ref{appendix:demonstration-exclusive}.


\subsection{Recovering the Nonautonomous Limit With a Special Observable of the Work Source}

The fluctuation theorems for nonautonomous exclusive work read $P^{\text{F}}(w_{\text{exc}}) / P^{\text{B}}(-w_{\text{exc}}) = e^{\beta w_{\text{exc}}} $ and $\langle e^{-\beta w_{\text{exc}}} \rangle_{\text{F}} = 1$~\cite{bochkovGeneralTheoryThermal1977,campisi_quantum_2011,campisiColloquiumQuantumFluctuation2011, jarzynskiFluctuationTheoremsAutonomous2025}. These relations are recovered from the fluctuation theorems for autonomous exclusive work when the term $- \text{ln}( p_{\xi^{\prime}}^{\text{tr}} / p_{\xi})$ in Eq.~\ref{SMeq:random-variable-rF} vanishes. This occurs when the measured observable $\hat{\Xi}_{R}$ of the work source commutes with its bare Hamiltonian $\hat{H}_{R}^{0}$ and the backaction of the system on the work source is negligible. Indeed, when $[\hat{\Xi}_{R},\hat{H}_{R}^{0}]=0$, the probability distribution of the measurement outcomes of $\hat{\Xi}_{R}$ remains unchanged under the evolution generated by $\hat{H}_{R}^{0}$. Consequently, in the absence of backaction, the additional logarithmic term in Eq.~\ref{SMeq:random-variable-rF} vanishes. However, if $\hat{\Xi}_{R}$ does not commute with $\hat{H}_{R}^{0}$, its probability distribution can evolve even in the absence of backaction. In this case, the fluctuation theorems for nonautonomous exclusive work cannot be recovered even in the limit of vanishing backaction.

\begin{figure}[!h]
    \centering
    \includegraphics{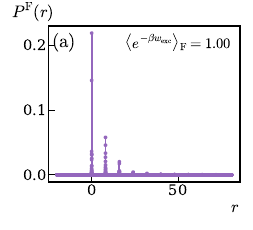}
    \includegraphics{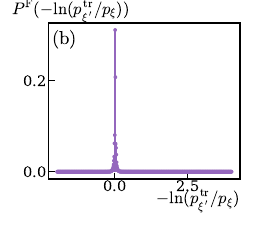}
    \caption{Recover the fluctuation theorems for nonautonomous exclusive work in the limit of vanishing backaction. (a) The distribution of the total entropy-like quantity $r$. (b) The distribution of the statistical change $-\ln(p_{\xi^{\prime}}^{\text{tr}}/p_{\xi})$ of the work source. In this plot, $\hbar=1,\beta=1,\omega_c = 8, \omega_z = \pi, \lambda = 0.6$. The number of atoms is $N = 400$. The dimension of the Hilbert space of the radiation field is truncated at $20$. The initial distribution of atoms on the eigenvalues of $\hat{J}_{z}$ is generated by the normal distribution with the average value $0$ and the standard deviation $50$ and then renormalized. The initial state of the radiation field is the thermal state of $\hat{H}_{S}^{0}$. The time of obtaining this plot is $\tau=5.2$.}
    \label{fig:exclusive-nonautonomous}
\end{figure}

We use the Dicke model to demonstrate the nonautonomous limit when the measured observable of the work source commutes with its bare Hamiltonian. We choose $\hat{\Xi}_{R}$ to be $\hat{J}_{z}$ of the ensemble of two-level atoms and the system energy is $\hat{H}_{S}^{0} = \hbar\omega_{c}\hat{a}^{\dagger}\hat{a}$. Figure~\ref{fig:exclusive-nonautonomous}(a) shows the distribution of the random variable $r$ in the limit of vanishing backaction. Figure~\ref{fig:exclusive-nonautonomous}(b) shows that, in the limit of vanishing backaction, the statistical change in the work source has a narrow distribution centered around zero. Hence, fluctuation theorems for the nonautonomous exclusive work can be recovered, which is numerically verified as $\langle e^{-\beta w_{\text{exc}}} \rangle_{\text{F}} = 1$ with the data shown in Fig.~\ref{fig:exclusive-nonautonomous}(a). 

\section{Discussion}
In this work, we extend fluctuation theorems for autonomous work from the classical regime to the quantum regime. 
To resolve the evolution between microscopic states in the composite Hilbert space, two-point measurements over both the work source and the system are adopted. For inclusive work, measuring the dynamical quantum work source determines not only the conditional energy spectrum of the system but also its fluctuating free-energy change. This yields fluctuation theorems in which work, conditional free energy, and work-source statistics appear as distinct contributions. 
The feasibility of the recovery of conventional nonautonomous fluctuation theorems depends qualitatively on whether work is defined inclusively or exclusively~\cite{jarzynskiComparisonFarfromequilibriumWork2007}. 
In the inclusive case, the autonomous fluctuation theorems cannot be consistently reduced to their nonautonomous counterparts~\cite{jarzynskiFluctuationTheoremsAutonomous2025} due to unavoidable spreading of the wave function of the observable of the work source. 
In the exclusive case, by contrast, the additional contribution associated with the work-source distribution vanishes when the backaction is negligible and the measured observable of the work source is conserved by its bare Hamiltonian. 
The established fluctuation theorems for autonomous work reveal the intrinsic fluctuation of the work source in the quantum regime.

We conclude with three remarks for future research. (i) Our analysis assumes a discrete observable of the work source to remain consistent with projective measurements. Extending the framework to continuous observables, e.g., the spatial coordinate, of the work source is an interesting problem, potentially requiring generalized measurements~\cite{watanabeGeneralizedEnergyMeasurements2014,micadeiQuantumFluctuationTheorems2020} and quasiprobability formulations of work~\cite{peiExploringQuasiprobabilityApproaches2023}. (ii) Fluctuation theorems have been established for classical nonautonomous and autonomous systems and quantum nonautonomous systems in the presence of heat baths, including regimes of weak and strong coupling between the system and the bath, provided that the thermodynamic quantities are defined appropriately~\cite{crooksEntropyProductionFluctuation1999,campisiColloquiumQuantumFluctuation2011,jarzynskiFluctuationTheoremsAutonomous2025}. Our framework for formulating fluctuation theorems for autonomous work in the quantum regime can be straightforwardly extended to the case in which the system is weakly coupled to a heat bath (see Appendix~\ref{appendix:weak-coupling-heat-bath} for details). Nevertheless, how to consistently extend the present framework to strong system-bath coupling in the quantum regime remains an open question. (iii) Applying the present framework to autonomous information and thermal machines~\cite{rasola_proposal_2025,mandalMaxwellsRefrigeratorExactly2013,chen_finite-time_2026}, as well as to information-processing systems involving measurement and feedback~\cite{shiraishi_role_2015,prech_quantum_2024}, may provide refined thermodynamic constraints on their performance and operation limits, particularly in microscopic regimes where fluctuations and the backaction associated with autonomous control may be significant.

\begin{acknowledgments}
Xiu-Hua Zhao would like to thank Ji-Hui Pei for helpful discussion. H. T. Quan acknowledges the support from the National Science Foundation of China under grants 12375028 and 12521004.
\end{acknowledgments}

\onecolumngrid{}

\appendix

\section{Detailed derivation and numerical methods about inclusive work}
\subsection{\label{appendix:time-reversal}The time-reversal process}
For a generic Hamiltonian $\hat{H}(t)$ governing the forward evolution from time $0$ to $\tau$, the corresponding Hamiltonian governing the time-reversal process is $\hat{H}^{\text{B}}(t) = \hat{\Theta}\hat{H}(\tau-t)\hat{\Theta}^{\dagger}$, where $\hat{\Theta}$ is the time-reversal operator~\cite{espositoNonequilibriumFluctuationsFluctuation2009,campisiColloquiumQuantumFluctuation2011}. While the forward evolution operator $\hat{U}(t,0)$ satisfies $d\hat{U}(t,0)/dt = -i\hbar^{-1}\hat{H}\hat{U}(t,0)$, the backward evolution operator $\hat{U}_{\text{B}}(t,0)$ satisfies $d\hat{U}_{\text{B}}(t,0)/dt = -i\hbar^{-1}\hat{H}^{\text{B}}\hat{U}_{\text{B}}(t,0)$. Using the formal solutions $\hat{U}(t,0) = \mathcal{T}\exp[-i\int_{0}^{t} dt^{\prime}\hat{H}(t^{\prime})/\hbar]$ and $\hat{U}_{\text{B}}(t,0) = \mathcal{T}\exp[-i\int_{0}^{t}dt^{\prime}\hat{H}^{\text{B}}(t^{\prime})/\hbar]$, we obtain
\begin{equation}
    \hat{U}_{\text{B}}(t, 0) = \hat{\Theta}\hat{U}(\tau-t, \tau)\hat{\Theta}^{\dagger},
\end{equation}
where $\mathcal{T}$ is the time-ordering operator. In our case, since the total Hamiltonian $\hat{H}_{\text{tot}}$ is time-independent, $\hat{U}_{\text{B}}(t, 0) = \hat{\Theta}\hat{U}(-t, 0)\hat{\Theta}^{\dagger}$.

In the forward process, we perform successive measurements over the variable $\hat{A}_{R}$ of the work source and the Hamiltonian $\hat{H}_{SR}$ at initial and final times, where the projector is expressed as $\hat{\Pi}^{a}_{n} = \ket{a, E_{n}^{a}}\bra{a, E_{n}^{a}}$. For the backward process, we consider the time-reversed projector $\hat{\tilde{\Pi}}^{a}_{n} = \hat{\Theta} \hat{\Pi}^{a}_{n}\hat{\Theta}^{\dagger}$, projecting onto the eigenstate $\hat{\Theta}\ket{a, E_{n}^{a}}$ of $\hat{\Theta}\hat{A}_{R}\hat{\Theta}^{\dagger}$ with the eigenvalue $a$ and $\hat{\Theta}\hat{H}_{S}(a)\hat{\Theta}^{\dagger}$ with the eigenvalue $E_n^{a}$. Here the eigenvalues remain the same as the forward process because they are real eigenvalues of Hermitian operators. In the backward process, the joint probability of two-point measurements, corresponding to the transition $\tilde{\Gamma}$ from $\hat{\Theta}|a^{\prime}, E_{m}^{a^{\prime}}\rangle$ to $\hat{\Theta}|a, E_{n}^{a}\rangle$, reads
\begin{equation}
\begin{aligned}
    P^{\text{B}}[\tilde{\Gamma}] =& \mathrm{Tr}\left\{ \hat{\tilde{\Pi}}_{n}^{a} \hat{U}_{\text{B}}(\tau,0) \hat{\tilde{\Pi}}_{m}^{a^{\prime}} \hat{\rho}_{0}^{\text{B}} \hat{\tilde{\Pi}}_{m}^{a^{\prime}} \hat{U}_{\text{B}}^{\dagger}(\tau,0) \hat{\tilde{\Pi}}_{n}^{a} \right\} \\
    =& \mathrm{Tr}\left\{ \hat{\Pi}_{n}^{a} \hat{U}(0,\tau) \hat{\Pi}_{m}^{a^{\prime}} \hat{\Theta}^{\dagger}\hat{\rho}_{0}^{\text{B}}\hat{\Theta} \hat{\Pi}_{m}^{a^{\prime}} \hat{U}^{\dagger}(0,\tau) \hat{\Pi}_{n}^{a} \right\},\\
    =& \mathrm{Tr}\left\{ \hat{\Pi}_{n}^{a} \hat{U}^{\dagger}(\tau,0) \hat{\Pi}_{m}^{a^{\prime}} \hat{\rho}_{0}^{\text{tr}} \hat{\Pi}_{m}^{a^{\prime}} \hat{U}(\tau,0) \hat{\Pi}_{n}^{a} \right\},
\end{aligned}
\end{equation}
where $\hat{\rho}_{0}^{\text{B}}$ is the initial state of the backward process. The second line is derived by inserting $\hat{\Theta}\hat{\Theta}^{\dagger}$ between each operator in the first line. The third line is obtained by defining $\hat{\rho}_{0}^{\text{tr}} \equiv \hat{\Theta}^{\dagger}\hat{\rho}_{0}^{\text{B}}\hat{\Theta}$ and using $\hat{U}(0,\tau)=\hat{U}^{-1}(\tau,0)=\hat{U}^{\dagger}(\tau,0)$.

\subsection{\label{appendix:numerical-method-Dicke-inclusive}Numerical representation of the Dicke model}

\begin{figure}[!h]
    \centering
    \includegraphics{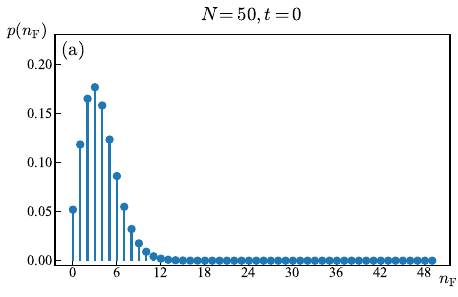}
    \includegraphics{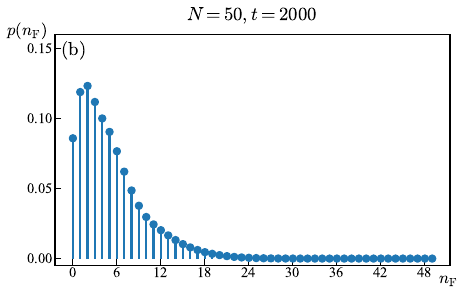}
    \includegraphics{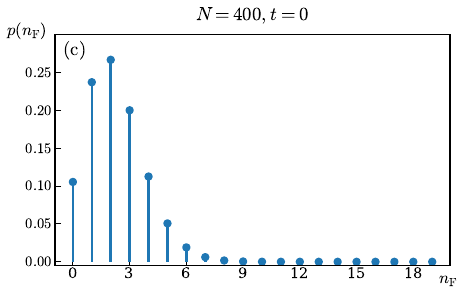}
    \includegraphics{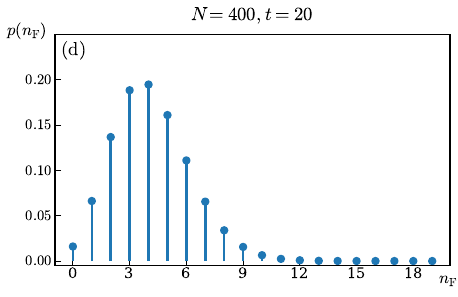}
    \caption{Convergence check of the truncated Fock states. (a) and (b), using the same parameters as those in Fig.~2(b-d) in the main text. $\hbar=1$, $\beta=3$, $\omega_c=0.8$, $\omega_z=0.4$, and $\lambda=0.25$. The number of atoms is $N = 50$. The dimension of the local Hilbert space of the radiation field is truncated at $50$. The initial distribution of atoms $p_{j}$ on the eigenvalues of $\hat{J}_{x}$ is generated by the normal distribution with the average value $-25$ and standard deviation $5$ and then renormalized. The initial state of the radiation field is the mixed thermal state for each $j$. (c) and (d), using the same parameters as those in Fig.~3 in the main text. $\hbar=1,\beta=1,\omega_c = 8, \omega_z = \pi, \lambda = 0.6$. The number of atoms is $N = 400$. The dimension of the Hilbert space of the radiation field is truncated at $20$. The initial state of the atoms is the eigenstate of $\hat{J}_{x}$ with the minimum eigenvalue, $j=-200$. The initial state of the radiation field is the thermal state with $j=-200$.}
    \label{SMfig:convergence_check_Fock_state}
\end{figure}

For the Hamiltonian $\hat{H}_{\text{tot}} = \hbar\omega_{c}\hat{I}_{\text{am}} \otimes \hat{a}^{\dagger}\hat{a} + \frac{2\lambda}{\sqrt{N}}\hat{J}_{x} \otimes \left(\hat{a}^{\dagger} + \hat{a}\right) + \omega_{z} \hat{J}_{z} \otimes \hat{I}_{\text{rf}}$, we perform numerical calculations with matrices of the operators expressed in the representation of the eigenstates of $\hat{J}_{x}$ of the atoms and the Fock states of the single-mode radiation field. This is convenient in the framework of inclusive work. 

In the local Hilbert space of the atoms and considering the subspace where the eigenvalue of $\hat{J}^2$ is $(N/2)(N/2+1)\hbar^2$, $\hat{J}_{x}$ is a diagonal matrix in its own representation, and the elements along the diagonal line are $(-N/2 + k)\hbar$ with $k=0, 1, 2, \dots, N$. To obtain the matrix of $\hat{J}_{z}$, we define the ladder operators $\hat{\mathcal{J}}_{\pm}^{x} \equiv \hat{J}_{y} \pm i \hat{J}_{z}$, which satisfy $[\hat{J}_{x}, \hat{\mathcal{J}}_{\pm}^{x}] = \pm \hbar \hat{\mathcal{J}}_{\pm}^{x}$ and $\hat{\mathcal{J}}_{-}^{x}\hat{\mathcal{J}}_{+}^{x} = \hat{J}^{2} - \hat{J}_{x}^{2} - \hbar\hat{J}_{x}$. $\hat{\mathcal{J}}_{+}^{x}$ is a matrix whose only non-zero entries lie on the first subdiagonal line, and the elements are chosen to  be $i\hbar\sqrt{(N/2)(N/2+1) - m(m+1)}$ with $m=-N/2, -N/2+1, -N/2+2, \dots, N/2-1$. $\hat{\mathcal{J}}_{-}^{x}$ is a matrix whose only non-zero entries lie on the first superdiagonal line and the elements are $-i\hbar\sqrt{(N/2)(N/2+1) - m(m-1)}$ with $m=-N/2+1, -N/2+2, \dots, N/2$. Consequently, $\hat{J}_{z} = (\hat{\mathcal{J}_{+}^{x}} -  \hat{\mathcal{J}_{-}^{x}})/(2i)$ is a real matrix. 

In the local Hilbert space of the radiation field, the creation operator $\hat{a}^{\dagger}$ is expressed by the matrix with nonzero elements $\sqrt{1}, \sqrt{2}, \sqrt{3}, \dots$ along the first subdiagonal line, and the annihilation operator $\hat{a}$ is expressed by the matrix with nonzero elements $\sqrt{1}, \sqrt{2}, \sqrt{3}, \dots$ along the first superdiagonal line. In the numerical calculation, the infinite-dimensional local Hilbert space of the radiation field is truncated at a finite dimension. The truncation dimension should be large enough to ensure that there is no excitation onto higher energy levels than the truncation value. Figure~\ref{SMfig:convergence_check_Fock_state} shows the distributions on the Fock states of the radiation field during the evolution. The parameters used in panels (a) and (b) are the same as those used in Fig.~2 in the main text. The parameters used in panels (c) and (d) are the same as those used in Fig.~3 in the main text. The excitations to high energy levels do not exceed the truncation dimension.


\subsection{\label{appendix:vanishing-backaction-Dicke}The limit of vanishing backaction in the Dicke model}
With the Hamiltonian $\hat{H}_{\text{tot}} = \hbar\omega_{c}\hat{I}_{\text{am}} \otimes \hat{a}^{\dagger}\hat{a} + \frac{2\lambda}{\sqrt{N}}\hat{J}_{x} \otimes \left(\hat{a}^{\dagger} + \hat{a}\right) + \omega_{z} \hat{J}_{z} \otimes \hat{I}_{\text{rf}}$, the second-order differential equation for the expectation value of $\hat{J}_{x}$ is
\begin{equation}
    \frac{d^2 \langle \hat{J}_{x} \rangle}{dt^2} = -
     \omega_z^2 \langle \hat{J}_{x} \rangle \left[ 1 + \delta_{\text{ba}} \right].
\end{equation}
Here, the parameter $\delta_{\text{ba}}$ characterizes the backaction of the radiation field on the atoms,
\begin{equation}
    \delta_{\text{ba}} \equiv - \frac{2\lambda}{\omega_{z}\sqrt{N}} \frac{\langle \hat{J}_{z} \otimes (\hat{a}^{\dagger} + \hat{a}) \rangle}{\langle \hat{J}_{x} \rangle}.
\end{equation}
The evolution of the radiation field satisfies
\begin{equation}
    \frac{d^2 \langle \hat{a}^{\dagger} + \hat{a} \rangle}{dt^2} = -\omega_{c}^2 \left[ \langle\hat{a}^{\dagger} + \hat{a}\rangle + \alpha_{\text{dr}} \right].
\end{equation}
Here, the parameter $\alpha_{\text{dr}}$ characterizes the driving of the atoms on the radiation field,
\begin{equation}
    \alpha_{\text{dr}} \equiv \frac{2\lambda}{\omega_{c}\sqrt{N}}\frac{2}{\hbar}\langle \hat{J}_{x} \rangle.
\end{equation}
When $\delta_{\text{ba}} \ll 1$ and $\alpha_{\text{dr}}\sim 1$, the backaction of the radiation field on the atoms can be neglected while the influence of the atoms on the radiation field is significant. Since the evolution of the atoms is weakly dependent on the radiation field in this limit, $\langle \hat{J}_{z} \otimes (\hat{a}^{\dagger} + \hat{a}) \rangle$ can be approximated by $\langle \hat{J}_{z} \rangle \langle (\hat{a}^{\dagger} + \hat{a}) \rangle$. Considering that the radiation field is strongly driven, we can make the following approximation $\langle \hat{a}^{\dagger} + a \rangle \sim -4\lambda\langle \hat{J}_{x} \rangle / (\hbar \omega_{c} \sqrt{N})$, which is the dragged center of the radiation field $\hbar\omega_{c}\hat{a}^{\dagger}\hat{a} + (2\lambda/\sqrt{N})\langle\hat{J}_{x}\rangle \left(\hat{a}^{\dagger} + \hat{a}\right)$. To determine appropriate parameters corresponding to a small $\delta_{\text{ba}}$, we consider the initial condition that $\langle \hat{J}_{z} \rangle \sim \hbar$. Therefore, $\delta_{\text{ba}} \sim \lambda^2 / (N \omega_{c} \omega_{z})$. Furthermore, with large expectation value of $\langle\hat{J}_{x}\rangle \sim N\hbar$, we have $\alpha_{\text{dr}} \sim \lambda\sqrt{N}/\omega_{c}$. 

\begin{figure}[!h]
    \centering
    \includegraphics[width=0.98\textwidth]{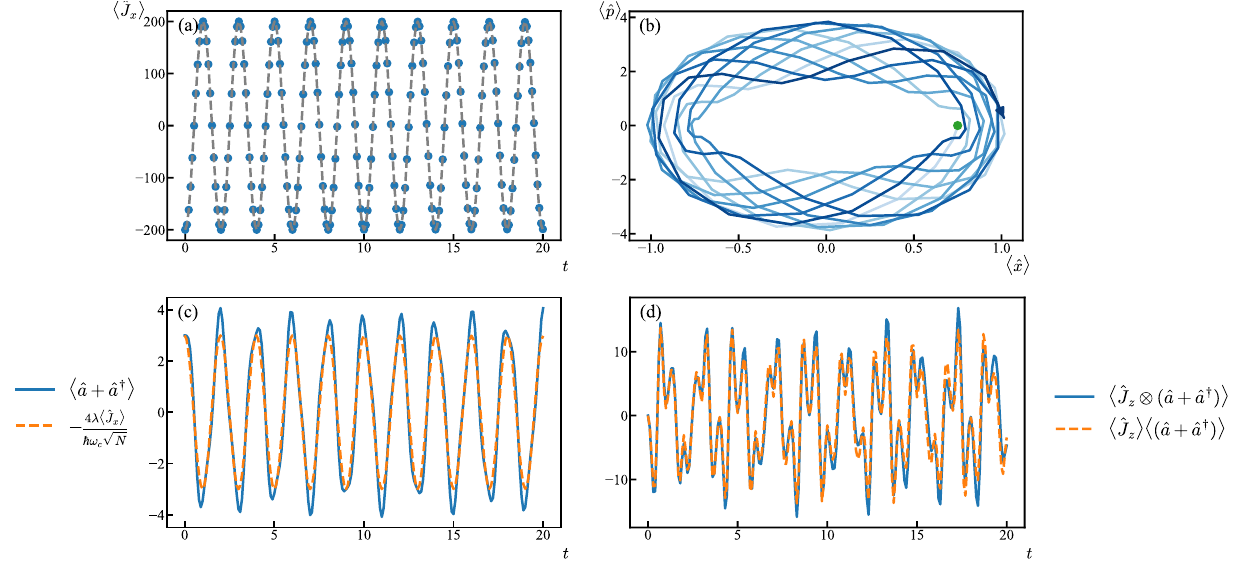}
    \caption{The evolution of the atoms and the radiation field in the regime of vanishing backaction. (a) The evolution of the expected value $\langle \hat{J}_{x} \rangle$. Blue data correspond to the evolution under the total Hamiltonian $\hat{H}_{\text{tot}}$, while the grey dashed line corresponds to $\langle\hat{J}_{x}\rangle_{t} = \langle\hat{J}_x\rangle_{0} \cos(\omega_z t) - \langle\hat{J}_{y}\rangle_{0} \sin(\omega_{z}t)$. (b) The phase-space trajectory of the radiation field under the total Hamiltonian $\hat{H}_{\text{tot}}$, where $\hat{x} \equiv \sqrt{\hbar / (2 m \omega_{c})} (\hat{a} + \hat{a}^{\dagger})$ and $\hat{p} \equiv - i \sqrt{\hbar m \omega_{c} / 2} (\hat{a} - \hat{a}^{\dagger})$ with $m=1$. The color indicates the time, from $t=0$ (light blue starting with a green circle marker) to $t=20$ (dark blue ending with an black arrow). (c) Comparing the expectation value of $\langle \hat{a}+\hat{a}^{\dagger}\rangle$ and the center of the oscillator dragged by the atoms. (d) Comparing the correlation and the factorized expectations between the atoms and the radiation field. Parameters used in this plot are the same as those in Fig.~3 in the main text. $\hbar=1,\beta=1,\omega_c = 8, \omega_z = \pi, \lambda = 0.6$. The number of atoms is $N = 400$. The dimension of the Hilbert space of the radiation field is truncated at $20$. The initial state of the ensemble of the atoms is the eigenstate of $\hat{J}_{x}$ with the minimum eigenvalue, $j=-200$. The initial state of the radiation field is the thermal state with $j=-200$.}
    \label{SMfig:vanishing_backaction_inclusive}
\end{figure}

Figure~\ref{SMfig:vanishing_backaction_inclusive} shows the evolution of the atoms and the radiation field when $\delta_{\text{ba}}\sim 3.58\times10^{-5}$ and $\alpha_{\text{dr}}\sim1.5$. The evolution of $\langle\hat{J}_{x}\rangle$ under the total Hamiltonian $\hat{H}_{\text{tot}}$ [blue data in Fig.~\ref{SMfig:vanishing_backaction_inclusive}(a)] is consistent with that under the bare Hamiltonian $\omega_{z}\hat{J}_{z}$ [gray dashed line in Fig.~\ref{SMfig:vanishing_backaction_inclusive}(a)]. The evolution of the radiation field in its phase space is significantly influenced by the atoms [Fig.~\ref{SMfig:vanishing_backaction_inclusive}(b)]. Figures~\ref{SMfig:vanishing_backaction_inclusive}(c) and (d) verify the approximation $\langle \hat{a}^{\dagger} + a \rangle \sim -4\lambda\langle \hat{J}_{x} \rangle / (\hbar \omega_{c} \sqrt{N})$ and $\langle \hat{J}_{z} \otimes (\hat{a}^{\dagger} + \hat{a}) \rangle \sim \langle \hat{J}_{z} \rangle \langle (\hat{a}^{\dagger} + \hat{a}) \rangle$, respectively.

\section{\label{appendix:factorized-semiclassical}The factorized semiclassical limit}

\subsection{General derivation}
With the Hamiltonian $\hat{H}_{\text{tot}} = \hat{I}_{R}\otimes\hat{H}_{S}^{0} + \hat{A}_{R}\otimes\hat{V}_{S} + \hat{H}_{R}^{0}\times\hat{I}_{S}$, the backaction of the system on the work source can be reflected on the second-order differential equation of $\langle\hat{A}_{R}\rangle$, that is, $\hbar^2 d^2\langle \hat{A}_{R} \rangle / dt^2 = \langle  [[\hat{H}_{R}^{0}, \hat{A}_{R}], \hat{H}_{R}^{0}]  \rangle + \langle  [[\hat{H}_{R}^{0}, \hat{A}_{R}], \hat{A}_{R}\otimes\hat{V}_{S} ]  \rangle $. We first consider the semi-decoupling limit where the backaction of the system on the work source can be neglected, and hence
\begin{equation}
\frac{d^2\langle \hat{A}_{R} \rangle }{dt^2} \approx \frac{1}{\hbar^2}\left\langle  \left[\left[\hat{H}_{R}^{0}, \hat{A}_{R}\right], \hat{H}_{R}^{0}\right]  \right\rangle.
\end{equation}
In this limit, the interaction term may be written as
\begin{equation}
    \hat{A}_{R}\otimes\hat{V}_{S} \equiv (\langle \hat{A}_{R} \rangle \hat{I}_{R} + \delta\hat{A}_{R})\otimes(\langle \hat{V}_{S} \rangle \hat{I}_{S} + \delta\hat{V}_{S})  \approx \langle \hat{V}_{S} \rangle \hat{A}_{R}\otimes\hat{I}_{S} + \langle \hat{A}_{R} \rangle \hat{I}_{R}\otimes\hat{V}_{S} - \langle \hat{A}_{R} \rangle \langle \hat{V}_{S} \rangle \hat{I}_{R}\otimes\hat{I}_{S},
\end{equation}
with $\delta \hat{A}_{R} \equiv \hat{A}_{R} - \langle \hat{A}_{R} \rangle \hat{I}_{R}$ and $\delta \hat{V}_{S} \equiv \hat{V}_{S} - \langle \hat{V}_{S} \rangle \hat{I}_{S}$. In the second approximate equality, we have neglected the correlation term $\delta \hat{A}_{R} \otimes \delta\hat{V}_{S}$, since the work source is weakly influenced by the system. Then the total Hamiltonian becomes
\begin{equation}
    \hat{H}_{\text{tot}}^{\text{sc}} \approx \hat{I}_{R} \otimes \hat{H}_{S}^{\text{sc}} + \hat{H}_{R}^{\text{sc}}\otimes\hat{I}_{S} - \langle \hat{A}_{R}\rangle\langle\hat{V}_{S}\rangle,
\end{equation}
where $\hat{H}_{S}^{\text{sc}} \equiv \hat{H}_{S}^{0} + \langle \hat{A}_{R} \rangle \hat{V}_{S}$, $\hat{H}_{R}^{\text{sc}} \equiv \hat{H}_{R}^{0} + \hat{A}_{R} \langle \hat{V}_{S} \rangle$, and again due to the limit of vanishing backaction, $\hat{H}_{R}^{\text{sc}} \approx \hat{H}_{R}^{0}$. The evolution operator of the composite system is also factorized as
\begin{equation}
    \hat{U}_{\text{sc}}(t,0) = e^{-i\hat{H}_{\text{tot}}t/\hbar} \approx e^{-i  \hat{H}_{R}^{0} t /\hbar} \otimes \mathcal{T}e^{-i \int_{0}^{t}\hat{H}_{S}^{\text{sc}} dt^{\prime}/\hbar} \equiv \hat{U}_{R}(t,0)\otimes\hat{U}_{S}(t,0),
\end{equation}
where the term $\langle \hat{A}_{R}\rangle\langle\hat{V}_{S}\rangle$ has been neglected since its contribution is a phase variation, and $\hat{U}_{\text{sc}}(t,0)$ and $\hat{U}_{\text{sc}}^{\dagger}(t,0)$ generally come in pairs in our derivation. 
Assume that the initial state can be factorized as
\begin{equation}
    \hat{\rho}_{\text{sc}0} = \hat{\rho}_{R0} \otimes\hat{\rho}_{S0}.
\end{equation}
Due to the factorized initial state and the factorized evolution operator, the evolution of the total density matrix $\hat{\rho}_{\text{sc}}$ satisfies $\hat{\rho}_{\text{sc}}(t) = \hat{\rho}_{R}(t)\otimes\hat{\rho}_{S}(t)$, where $\hat{\rho}_{R}(t) = \hat{U}_{R}(t,0)\hat{\rho}_{R0}\hat{U}_{R}^{\dagger}(t,0)$ and the evolution of the system is governed by
\begin{equation}
    i\hbar\frac{d\hat{\rho}_{S}}{dt} = \left[\hat{H}_{S}^{0} + \langle \hat{A}_{R} \rangle \hat{V}_{S} , \hat{\rho}_{S}\right].
\end{equation}
The above equation resembles the evolution of a system with the time-dependent Hamiltonian $\hat{H}_{S}^{\text{sc}}(\langle \hat{A}_{R} \rangle_{t}) \equiv \hat{H}_{S}^{0} + \langle \hat{A}_{R} \rangle_{t} \hat{V}_{S}$. Then the fluctuation theorem for a driven Hamiltonian can be recovered with the following initial state
\begin{equation}
    \hat{\rho}_{R0} = \hat{\rho}_{R0} \; (\text{arbitrary}), \quad \hat{\rho}_{S0} = e^{-\beta \hat{H}_{S}^{\text{sc}}(\langle \hat{A}_{R} \rangle_{0})} / Z_{S}^{\text{sc}}(\langle \hat{A}_{R} \rangle_{0}),
\end{equation}
where $Z_{S}^{\text{sc}}(\langle \hat{A}_{R} \rangle_{t}) \equiv \mathrm{Tr}_{S}\{e^{-\beta \hat{H}_{S}^{\text{sc}}(\langle \hat{A}_{R} \rangle_{t})}\}$. The joint projectors in the autonomous scenario [Eq.~5 in the main text] are replaced by
\begin{equation}
    \hat{B}_{n}^{0} = \hat{I}_{R} \otimes \ket{\epsilon_{n}^{0}}\bra{\epsilon_{n}^{0}},
\end{equation}
and 
\begin{equation}
    \hat{B}_{m}^{\tau} = \hat{I}_{R} \otimes \ket{\epsilon_{m}^{\tau}}\bra{\epsilon_{m}^{\tau}},
\end{equation}
for initial and final measurements, respectively. 
Here, $\ket{\epsilon_{n}^{0}}$ and $\ket{\epsilon_{m}^{\tau}}$ are respectively the eigenstates of $\hat{H}_{S}^{\text{sc}}(\langle \hat{A}_{R} \rangle_{0})$ and $\hat{H}_{S}^{\text{sc}}(\langle \hat{A}_{R} \rangle_{\tau})$. 
The probability of initially sampling $\epsilon_{n}^{0}$ and finally sampling $\epsilon_{m}^{\tau}$ in the forward process is
\begin{equation}
\begin{aligned}
    P^{\text{F}}[m\gets n] \equiv& \mathrm{Tr}\left\{ \hat{B}_{m}^{\tau} \hat{U}_{\text{sc}}(\tau,0) \hat{B}_{n}^{0} \hat{\rho}_{\text{sc}0} \hat{B}_{n}^{0} \hat{U}_{\text{sc}}^{\dagger}(\tau,0) \hat{B}_{m}^{\tau} \right\} \\
    =& \mathrm{Tr}_{R}\left\{ \hat{U}_{R}(\tau,0)\hat{\rho}_{R0} \hat{U}_{R}^{\dagger}(\tau,0)\right\} \mathrm{{Tr}}_{S} \left\{ \ket{\epsilon_{m}^{\tau}}\bra{\epsilon_{m}^{\tau}} \hat{U}_{S}(\tau,0) \ket{\epsilon_{n}^{0}}\bra{\epsilon_{n}^{0}} \hat{\rho}_{S0} \ket{\epsilon_{n}^{0}}\bra{\epsilon_{n}^{0}} \hat{U}_{S}^{\dagger}(\tau,0) \ket{\epsilon_{m}^{\tau}}\bra{\epsilon_{m}^{\tau}} \right\} \\
    =& \left| \bra{\epsilon_{m}^{\tau}}\hat{U}_{S}(\tau,0)\ket{\epsilon_{n}^{0}} \right|^2 \bra{\epsilon_{n}^{0}} \hat{\rho}_{S0}\ket{\epsilon_{n}^{0}}.
\end{aligned}
\end{equation}
In the time-reversed process, the evolution operator from time $0$ to $t$ becomes $\hat{\Theta}\hat{U}_{\text{sc}}(\tau-t,\tau)\hat{\Theta}^{\dagger}$. The projectors should also be reversed. Starting from the time reversal of the final state of the work source in the forward process, the evolution of $\langle \hat{A}_{R} \rangle _t$ is reversed in the backward process. From an initial state $\hat{\rho}_{\text{sc}0}^{\text{B}}$ of the composite system, the probability of initially projecting the system to $\epsilon_{m}^{\tau}$ and finally projecting the system to $\epsilon_{n}^{0}$ in the backward process is finally expressed as
\begin{equation}
\begin{aligned}
    P^{\text{B}}[n\gets m] \equiv& \mathrm{Tr}\left\{ \hat{\Theta} \hat{B}_{n}^{0} \hat{\Theta}^{\dagger} \hat{\Theta}\hat{U}_{\text{sc}}(0,\tau)\hat{\Theta}^{\dagger} \hat{\Theta} \hat{B}_{m}^{\tau} \hat{\Theta}^{\dagger} \hat{\rho}_{\text{sc}0}^{\text{B}} \hat{\Theta} \hat{B}_{m}^{\tau} \hat{\Theta}^{\dagger} \hat{\Theta} \hat{U}_{\text{sc}}^{\dagger}(0,\tau) \hat{\Theta}^{\dagger} \hat{\Theta} \hat{B}_{n}^{0} \hat{\Theta}^{\dagger} \right\} \\
    =&\mathrm{Tr}\left\{ \hat{B}_{n}^{0} \hat{U}_{\text{sc}}^{\dagger}(\tau,0) \hat{B}_{m}^{\tau} \hat{\rho}_{\text{sc}0}^{\text{tr}} \hat{B}_{m}^{\tau} \hat{U}_{\text{sc}}(\tau,0) \hat{B}_{n}^{0} \right\} \\
    =& \mathrm{Tr}_{R}\left\{ \hat{U}^{\dagger}_{R}(\tau,0)\hat{\rho}_{R0}^{\text{tr}} \hat{U}_{R}(\tau,0)\right\} \mathrm{{Tr}}_{S} \left\{ \ket{\epsilon_{n}^{0}}\bra{\epsilon_{n}^{0}} \hat{U}^{\dagger}_{S}(\tau,0) \ket{\epsilon_{m}^{\tau}}\bra{\epsilon_{m}^{\tau}} \hat{\rho}_{S0}^{\text{tr}} \ket{\epsilon_{m}^{\tau}}\bra{\epsilon_{m}^{\tau}} \hat{U}_{S}(\tau,0)\right\} \\
    =& \left| \bra{\epsilon_{m}^{\tau}}\hat{U}_{S}(\tau,0)\ket{\epsilon_{n}^{0}} \right|^2 \bra{\epsilon_{m}^{\tau}} \hat{\rho}_{S0}^{\text{tr}}\ket{\epsilon_{m}^{\tau}},
\end{aligned}
\end{equation}
where
\begin{equation}
    \hat{\rho}_{R0}^{\text{tr}} = \hat{U}_{R}(\tau,0)\hat{\rho}_{R0}\hat{U}_{R}^{\dagger}(\tau,0), \quad \hat{\rho}_{S0}^{\text{tr}} = e^{-\beta \hat{H}_{S}^{\text{sc}}(\langle \hat{A}_{R} \rangle_{\tau})} / Z_{S}^{\text{sc}}(\langle \hat{A}_{R} \rangle_{\tau}).
\end{equation}
We have defined $\hat{\rho}_{\text{sc}0}^{\text{tr}} = \hat{\Theta}^{\dagger}\hat{\rho}_{\text{sc}0}^{\text{B}}\hat{\Theta}$ and $\hat{\rho}_{\text{sc}0}^{\text{tr}} = \hat{\rho}_{R0}^{\text{tr}}\otimes\hat{\rho}_{S0}^{\text{tr}}$.

The work performed on the system in the forward and backward processes are respectively defined as
\begin{align}
    &w_{\text{sc}}[m\gets n] \equiv \epsilon_{m}^{\tau} - \epsilon_{n}^{0} = \frac{1}{\beta}\ln\frac{P^{\text{F}}[m\gets n]}{P^{\text{B}}[n\gets m]} + \Delta F_{S}^{\text{sc}}, \\
    &w_{\text{sc}}[n \gets m] \equiv \epsilon_{n}^{0} - \epsilon_{m}^{\tau} = \frac{1}{\beta}\ln\frac{P^{\text{B}}[n\gets m]}{P^{\text{F}}[m\gets n]} - \Delta F_{S}^{\text{sc}},
\end{align}
where 
\begin{equation}
    \Delta F_{S}^{\text{sc}} \equiv -\frac{1}{\beta} \ln Z_{S}^{\text{sc}}(\langle \hat{A}_{R} \rangle_{\tau}) + \frac{1}{\beta} \ln Z_{S}^{\text{sc}}(\langle \hat{A}_{R} \rangle_{0}).
\end{equation}
is the free energy change from $t=0$ to $t=\tau$ in the forward process. 
Considering the probability distribution of $ w_{\text{sc}}[m\gets n]$ in the forward process and $w_{\text{sc}}[n \gets m]$ in the backward process, 
\begin{align}
    &P^{\text{F}}(w_{\text{sc}}) \equiv \sum_{m,n} P^{\text{F}}[m\gets n]\delta\left( w_{\text{sc}}[m\gets n] - w_{\text{sc}} \right), \\ &P^{\text{B}}(w_{\text{sc}}) \equiv \sum_{n,m} P^{\text{B}}[n \gets m]\delta\left( w_{\text{sc}}[n \gets m] - w_{\text{sc}} \right),
\end{align}
we obtain the fluctuation theorems for work in the factorized semiclassical limit,
\begin{equation}
    \frac{P^{\text{F}}(w_{\text{sc}})}{P^{\text{B}}(-w_{\text{sc}})} = e^{\beta (w_{\text{sc}} - \Delta F_{S}^{\text{sc}})} ,
\end{equation}
and
\begin{equation}
    \langle e^{-\beta w_{\text{sc}}} \rangle_{\text{F}} = e^{-\beta \Delta F_{S}^{\text{sc}}}.
\end{equation}

\subsection{Factorized semiclassical limit in the Dicke model}
We use the Dicke model to demonstrate the evolution in the factorized semiclassical limit, which is tied to the limit of vanishing backaction. With the Hamiltonian $\hat{H}_{\text{tot}} = \hbar\omega_{c}\hat{I}_{\text{am}} \otimes \hat{a}^{\dagger}\hat{a} + \frac{2\lambda}{\sqrt{N}}\hat{J}_{x} \otimes \left(\hat{a}^{\dagger} + \hat{a}\right) + \omega_{z} \hat{J}_{z} \otimes \hat{I}_{\text{rf}}$, in the semi-deoupling limit where the backaction of the radiation field on the atoms is negligible, the evolution of the atoms is governed by
\begin{equation}
    \hat{U}_{R}(t, 0) = \exp\left\{- \frac{i}{\hbar}\omega_{z} \hat{J}_{z} t \right\},
\end{equation}
and
\begin{equation}
    \langle \hat{J}_{x} \rangle _{t} = \langle \hat{J}_{x} \rangle _{0} \cos \omega_{z}t - \langle \hat{J}_{y} \rangle _{0} \sin \omega_{z}t.
\end{equation}
Following the procedure of semiclassical factorization in the last section, the effective Hamiltonian of the radiation field is $\hat{H}_{S}^{\text{sc}}(\langle\hat{J}_{x}\rangle_{t}) = \hbar\omega_{c}\hat{a}^{\dagger}\hat{a} + \frac{2\lambda}{\sqrt{N}}\langle\hat{J}_{x}\rangle_{t} (\hat{a}^{\dagger} + \hat{a})$, with the evolution operator,
\begin{equation}
    \hat{U}_{S}(t,0) = \mathcal{T} \exp\left\{ - \frac{i}{\hbar} \int_{0}^{t} \hat{H}_{S}^{\text{sc}}(\langle\hat{J}_{x}\rangle_{t^{\prime}})  dt^{\prime}\right\} = \mathcal{T} \exp\left\{ - \frac{i}{\hbar} \int_{0}^{t} \left[\hbar\omega_{c}\hat{a}^{\dagger}\hat{a} + \frac{2\lambda}{\sqrt{N}}\langle\hat{J}_{x}\rangle_{t^{\prime}} (\hat{a}^{\dagger} + \hat{a}) \right] dt^{\prime}\right\}.
\end{equation}
The explicit expression of $\hat{U}_{S}(t,0)$ can be obtained in the interaction picture. The relation between states of the radiation field $\hat{\rho}_{S}(t)$ in the Schrodinger picture and $\hat{\rho}_{S}^{I}(t)$ in the interaction picture is $\hat{\rho}_{S}^{I}(t) = \hat{U}_{0}^{\dagger}(t,0)\hat{\rho}_{S}(t)\hat{U}_{0}(t,0)$, where $\hat{U}_{0}(t,0) = \exp \left\{ - i \omega_{c}\hat{a}^{\dagger}\hat{a}t  \right\}$. The evolution operator in the interaction picture is
\begin{equation}
\begin{aligned}
    \hat{U}_{S}^{I}(t, 0) =& \mathcal{T} \exp\left\{ - \frac{i}{\hbar} \int_{0}^{t} \left[ \hat{U}_{0}^{\dagger}(t^{\prime},0) \frac{2\lambda}{\sqrt{N}}\langle\hat{J}_{x}\rangle_{t^{\prime}} (\hat{a}^{\dagger} + \hat{a}) \hat{U}_{0}(t^{\prime},0)dt^{\prime}\right]\right\} \\
    =&\mathcal{T} \exp\left\{ - \frac{i}{\hbar} \int_{0}^{t} \left[ \frac{2\lambda}{\sqrt{N}}\langle\hat{J}_{x}\rangle_{t^{\prime}} e^{ i \omega_{c}\hat{a}^{\dagger}\hat{a} t^{\prime}}  (\hat{a}^{\dagger} + \hat{a}) e^{ - i \omega_{c}\hat{a}^{\dagger}\hat{a} t^{\prime}} dt^{\prime}\right]\right\} \\
    =& \mathcal{T} \exp\left\{ - \frac{i}{\hbar} \int_{0}^{t} \left[ \frac{2\lambda}{\sqrt{N}}\langle\hat{J}_{x}\rangle_{t^{\prime}} \left( e^{i\omega_{c}t^{\prime}} \hat{a}^{\dagger} + e^{-i\omega_{c}t^{\prime}}\hat{a} \right)  dt^{\prime}\right]\right\}.
\end{aligned}
\end{equation}
Denoting $\hat{H}^{I}_{S}(t^{\prime}) \equiv \frac{2\lambda}{\sqrt{N}}\langle\hat{J}_{x}\rangle_{t^{\prime}} \left( e^{i\omega_{c}t^{\prime}} \hat{a}^{\dagger} + e^{-i\omega_{c}t^{\prime}}\hat{a} \right)  $, so that $\hat{U}_{S}^{I}(t, 0) = \mathcal{T} \exp[\int_{0}^{t} - i \hbar^{-1} \hat{H}_{S}^{I}(t^{\prime}) dt^{\prime}]$. Applying the Magnus expansion~\cite{blanes_magnus_2009}, we obtain
\begin{equation}
    \hat{U}_{S}^{I}(t, 0) = e^{\hat{\Omega}_{1}(t ,0) + \Omega_{2}(t,0)}
\end{equation}
where
\begin{equation}
    \hat{\Omega}_{1}(t,0) \equiv - \frac{i}{\hbar}\int_{0}^{t} \hat{H}_{S}^{I}(t_{1}) dt_{1} = -\frac{i}{\hbar} \frac{2\lambda}{\sqrt{N}} \hat{a}^{\dagger} \int_{0}^{t} \langle \hat{J}_{x} \rangle _{t_{1}} e^{i\omega_{c}t_{1}}dt_{1} - \frac{i}{\hbar} \frac{2\lambda}{\sqrt{N}} \hat{a} \int_{0}^{t} \langle \hat{J}_{x} \rangle _{t_{1}} e^{-i\omega_{c}t_{1}}dt_{1} = \alpha \hat{a}^{\dagger} -  \alpha^{*} \hat{a},
\end{equation}
\begin{equation}
    \alpha = -\frac{i}{\hbar} \frac{2\lambda}{\sqrt{N}} \frac{-e^{i \omega_{c} t} \left[\sin (\omega_{z} t) \left(\langle \hat{J}_{x} \rangle _{0} \omega_{z}-i \langle \hat{J}_{y} \rangle _{0} \omega_{c}\right)+\cos (\omega_{z} t) \left(\langle \hat{J}_{y} \rangle _{0} \omega_{z}+i \langle \hat{J}_{x} \rangle _{0} \omega_{c}\right) \right]+i \langle \hat{J}_{x} \rangle _{0} \omega_{c}+\langle \hat{J}_{y} \rangle _{0} \omega_{z}}{(\omega_{c}-\omega_{z}) (\omega_{c}+\omega_{z})},
\end{equation}
and
\begin{equation}
\begin{aligned}
    \Omega_{2}(t,0) \equiv& -\frac{1}{2\hbar^2} \int_{0}^{t}dt_{1} \int_{0}^{t_{1}}dt_{2} [\hat{H}_{S}^{I}(t_{1}), \hat{H}_{S}^{I}(t_{2})] \\
    =& \frac{1}{\hbar^2} \frac{2\lambda^2}{N} \int_{0}^{t}dt_{1} \int_{0}^{t_{1}}dt_{2} \langle \hat{J}_{x} \rangle _{t_{1}} \langle \hat{J}_{x} \rangle _{t_{2}} e^{i\omega_{c}t_{1}} e^{-i\omega_{c}t_{2}} - \frac{1}{\hbar^2} \frac{2\lambda^2}{N} \int_{0}^{t}dt_{1} \int_{0}^{t_{1}}dt_{2} \langle \hat{J}_{x} \rangle _{t_{1}} \langle \hat{J}_{x} \rangle _{t_{2}} e^{- i\omega_{c}t_{1}} e^{i\omega_{c}t_{2}}.
\end{aligned}
\end{equation}
Higher order terms in the Magnus expansion vanish since the commutators between $\hat{H}_{S}^{I}(t)$ at different time are c-numbers. The evolution operator for the radiation field in the Schrodinger picture reads
\begin{equation}
    \hat{U}_{S}(t, 0) = \hat{U}_{0}(t,0)\hat{U}_{S}^{I}(t, 0), \quad \hat{U}_{S}^{I}(t, 0) = e^{\alpha \hat{a}^{\dagger}-\alpha^* \hat{a}},
\end{equation}
where we have neglected the pure phase term, since $\hat{U}_{S}(t,0)$ and $\hat{U}_{S}^{\dagger}(t,0)$ generally come in pairs in our derivation.

\begin{figure}[!h]
    \centering
    \includegraphics{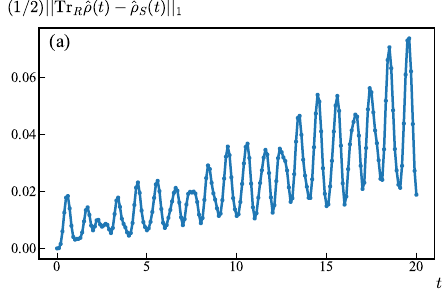}
    \includegraphics{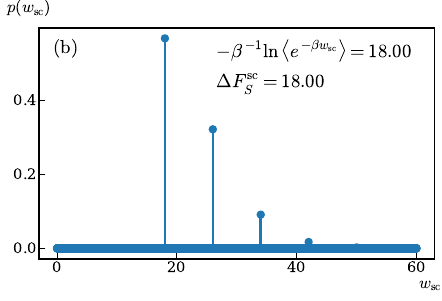}
    \caption{Factorized semiclassical limit of the Dicke model. (a) The trace distance between two states. The first one is the state of the radiation field under the total Hamiltonian, which is given by tracing out the atoms from the total density matrix, that is, $\mathrm{Tr}_{R}\hat{\rho}(t)$. The second one is the state of the radiation field $\hat{\rho}_{S}(t)$ under the effective driven Hamiltonian. (b) The work distribution obtained by the projectors $\ket{\epsilon_{n}^{0}}\bra{\epsilon_{n}^{0}}$ and $\ket{\epsilon_{m}^{\tau}}\bra{\epsilon_{m}^{\tau}}$ on the radiation field driven by the effective Hamiltonian $\hat{H}_{S}^{\text{sc}}(\langle\hat{J}_{x}\rangle_{t})$ at $\tau=5.5$. The free energy difference is obtained by $\Delta F_{S}^{\text{sc}} = F_{S}^{\text{sc}}(\langle\hat{J}_{x}\rangle_{\tau}) - F_{S}^{\text{sc}}(\langle\hat{J}_{x}\rangle_{0})$ with $F_{S}^{\text{sc}}(\langle\hat{J}_{x}\rangle_{t}) = -\beta^{-1}\ln \mathrm{Tr}_{S}[e^{-\beta \hat{H}_{S}^{\text{sc}}(\langle\hat{J}_{x}\rangle_t)}]$. In this plot, the parameters and initial conditions are the same as those in Fig.~\ref{SMfig:vanishing_backaction_inclusive}.}
    \label{SMfig:semiclassical_meanfield_inclusive}
\end{figure}

In the parameter regime of semi-decoupling, we compare the evolutions of the radiation field under the total Hamiltonian $\hat{H}_{\text{tot}}$ with the evolution operator $\hat{U}(t,0)$ and the effective time-dependent Hamiltonian $\hat{H}_{S}^{\text{sc}}(\langle\hat{J}_{x}\rangle_{t})$ with the effective evolution operator $\hat{U}_{S}(t,0)$. Figure~\ref{SMfig:semiclassical_meanfield_inclusive}(a) shows the evolution of the trace distance between the states of the radiation field under total and effective dynamics. $\mathrm{Tr}_{R}\hat{\rho}(t)$ represents the state of the radiation field obtained by tracing out the atoms in the composite state $\hat{\rho}(t)$, which is governed by the total evolution operator $\hat{U}(t,0)$. $\hat{\rho}_{S}(t)$ is governed by the effective evolution $\hat{U}_{S}(t,0)$. The small value of the trace distance $(1/2)||\mathrm{Tr}_{R}\hat{\rho}-\hat{\rho}_{S}||_{1}=(1/2)\mathrm{Tr}_{S}\sqrt{(\mathrm{Tr}_{R}\hat{\rho}-\hat{\rho}_{S})^\dagger(\mathrm{Tr}_{R}\hat{\rho}-\hat{\rho}_{S})}$ between $\hat{\rho}$ and $\hat{\rho}_{S}$ at early times indicates that the factorized
semiclassical approximation accurately describes the early-stage
evolution. Figure~\ref{SMfig:semiclassical_meanfield_inclusive}(b) shows the work statistics under the time-dependent effective Hamiltonian and projectors, demonstrating the nonautonomous fluctuation theorems for work in the semiclassical approximation.

\section{\label{appendix:demonstration-exclusive}Demonstration of fluctuation theorems for exclusive work}
We use the Dicke model $\hat{H}_{\text{tot}} = \hbar\omega_{c}\hat{I}_{\text{am}} \otimes \hat{a}^{\dagger}\hat{a} + \frac{2\lambda}{\sqrt{N}}\hat{J}_{x} \otimes \left(\hat{a}^{\dagger} + \hat{a}\right) + \omega_{z} \hat{J}_{z} \otimes \hat{I}_{\text{rf}}$ to demonstrate the fluctuation theorems for exclusive work. In the local Hilbert space of the atoms, considering the subspace where the eigenvalue of $\hat{J}^2$ is $(N/2)(N/2+1)\hbar^2$, the matrices of operators are expressed with the representation of $\hat{J}_{z}$. Then $\hat{J}_{z}$ is a diagonal matrix and the elements along the diagonal line are $(-N/2 + k)\hbar$ with $k=0, 1, 2, \dots, N$. With the help of ladder operators, $\hat{\mathcal{J}}_{\pm}^{z} \equiv \hat{J}_{x} \pm i \hat{J}_{y}$, satisfying $[\hat{J}_{z}, \hat{\mathcal{J}_{\pm}^{z}}] = \pm \hbar\hat{\mathcal{J}}_{\pm}^{z}$ and $\hat{\mathcal{J}}_{-}^{z}\hat{\mathcal{J}}_{+}^{z} = \hat{J}^2 - \hat{J}_{z}^{2} - \hbar \hat{J}_{z}$, we can obtain the matrix of $\hat{J}_{x}$. $\hat{\mathcal{J}}_{+}^{z}$ is a matrix whose only non-zero entries lie on the first subdiagonal line, and the elements are $\sqrt{(N/2)(N/2+1) - m(m+1)}\hbar$ with $m=-N/2, -N/2+1, -N/2+2, \dots, N/2-1$. $\hat{\mathcal{J}}_{-}^{z}$ is a matrix whose only non-zero entries lie on the first superdiagonal line and the elements are $\sqrt{(N/2)(N/2+1) - m(m-1)}\hbar$ with $m=-N/2+1, -N/2+2, \dots, N/2$. Then $\hat{J}_{x} = ( \hat{\mathcal{J}_{+}^{z}} +  \hat{\mathcal{J}_{-}^{z}})/2$. In the local Hilbert space of the radiation field, the operators are expressed with the representation of the Fock states. 

\begin{figure}[!h]
    \centering
    \includegraphics{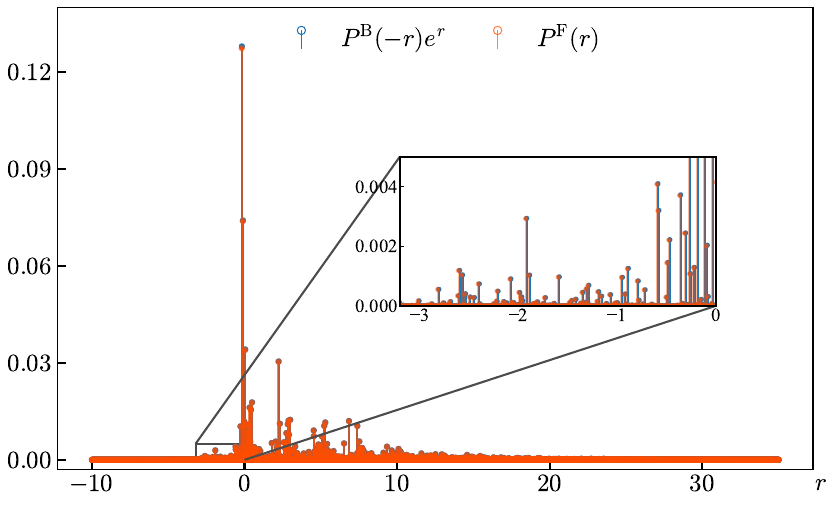}
    \caption{Verification of the fluctuation theorems for the autonomous exclusive work. $\hbar=1$, $\beta=3$, $\omega_c=0.8$,$\omega_z=0.4$,$\lambda=0.25$. The number of atoms is $N = 50$. The dimension of the local Hilbert space of the radiation field is truncated at $50$. The initial distribution of atoms on the eigenvalues of $\hat{J}_{z}$ in the forward process is generated by the normal distribution with the average value $-25$ and standard deviation $5$ and then renormalized. The time of obtaining this plot is $\tau=1000$.}
    \label{SMfig:exclusive_statistics}
\end{figure}

To verify the fluctuation theorems, we perform joint measurements over $\hat{J}_{z}$ and $\hat{H}_{S}^{0} = \hbar\omega_{c}\hat{a}^{\dagger}\hat{a}$. Figure~\ref{SMfig:exclusive_statistics} verifies the fluctuation theorems for the autonomous exclusive work in a general case. 

\section{\label{appendix:weak-coupling-heat-bath}Weak coupling to a heat bath}

We now extend our framework of fluctuation theorems for autonomous work in the quantum regime to a system ($S$) weakly coupled to a heat bath ($B$). The work source ($R$) does not interact directly with the bath. The total Hamiltonian of the work source, the system, and the bath reads
\begin{equation}
    \hat{H}_{\text{tot}}^{RSB} = \hat{H}_{R}^{0}\otimes\hat{I}_{S} \otimes \hat{I}_{B} + \hat{A}_{R}\otimes\hat{V}_{S}\otimes\hat{I}_{B} + \hat{I}_{R} \otimes \hat{H}_{S}^{0} \otimes \hat{I}_{B} + \hat{I}_{R} \otimes \hat{H}_{SB} + \hat{I}_{R}\otimes\hat{I}_{S}\otimes\hat{H}_{B}^{0},
\end{equation}
where $\hat{H}_{R}^{0}$, $\hat{H}_{S}^{0}$, and $\hat{H}_{B}^{0}$ are the bare Hamiltonians of the work source, the system, and the heat bath, respectively. $\hat{A}_{R}\otimes\hat{V}_{S}$ describes the interaction between the work source and the system, while $\hat{H}_{SB}$ describes the interaction between the system and the bath. The operators $\hat{I}_{R}$, $\hat{I}_{S}$, and $\hat{I}_{B}$ denote the identity operators on the corresponding local Hilbert spaces.
We assume that the system-bath coupling is sufficiently weak that the system-bath interaction energy, as well as its change between the initial and final times, can be neglected compared with the relevant changes in the system and bath energies. Under this approximation, the negative change in the bare bath energy can be identified with the heat absorbed by the system~\cite{campisiColloquiumQuantumFluctuation2011,deffner_nonequilibrium_2011}. We also assume a factorized initial preparation between the composite work-source-system subsystem and the bath, thereby neglecting initial system-bath correlations.
During the evolution, the composite system consisting of the work source, the system, and the bath is treated as a time-independent isolated quantum system. The forward evolution operator is denoted by $\hat{U}_{RSB}(t,0)=\exp(-i\hat{H}_{\text{tot}}t/\hbar)$. The corresponding backward evolution operator from time $0$ to $t$ reads $\hat{\Theta}\hat{U}_{RSB}(-t, 0)\hat{\Theta}^{\dagger}$, since the total Hamiltonian is time-independent.

\subsection{Inclusive work}
In the absence of the bath, the framework of formulating fluctuation theorems for the inclusive work consists of successive projective measurements over the observable $\hat{A}_{R}$ of the work source and the conditional Hamiltonian $\hat{H}_{S}(a) = \hat{H}_{S}^{0} + a\hat{V}_{S}$ of the system at initial and final times. In the presence of the bath, we additionally measure the bare Hamiltonian $\hat{H}_{B}^{0}$ of the bath. Denoting the eigenvalue of $\hat{H}_{B}^{0}$ as $E_{\mu}^{B}$ and following the similar procedure as in the case without a heat bath, we obtain the joint probability of the outcome of the initial measurements being $|a,E_{n}^{a},E_{\mu}^{B}\rangle$ and the outcome of the final measurements being $|a^{\prime},E_{m}^{a^{\prime}},E_{\nu}^{B}\rangle$ as
\begin{equation}
    P^{\text{F}}[\Gamma] = \left| \langle a^{\prime}, E_{m}^{a^{\prime}}, E_{\nu}^{B} | \hat{U}_{RSB} (\tau, 0) | a, E_{n}^{a}, E_{\mu}^{B} \rangle \right|^2 \langle a, E_{n}^{a}, E_{\mu}^{B} | \hat{\rho}_{RSB} | a, E_{n}^{a}, E_{\mu}^{B} \rangle,
\end{equation}
where $\Gamma$ denotes the transition from $|a,E_{n}^{a},E_{\mu}^{B}\rangle$ to $|a^{\prime},E_{m}^{a^{\prime}},E_{\nu}^{B}\rangle$ in the forward process. For notational simplicity, we write the above expression with a nondegenerate bath spectrum. Degeneracies can be treated by replacing the single-valued projectors with the corresponding spectral projectors~\cite{espositoNonequilibriumFluctuationsFluctuation2009}, without changing the form of the following fluctuation theorems. 
The initial state of the total system in the forward process is
\begin{equation}
    \hat{\rho}_{RSB} = \left[ \sum_{a} p_{a} \ket{a}\bra{a} \otimes e^{-\beta \hat{H}_{S}(a)} Z_{S}^{-1}(a) \right] \otimes e^{-\beta \hat{H}_{B}^{0}} Z_{B}^{-1},
\end{equation}
where $Z_{S}(a) = \mathrm{Tr}_{S}[e^{-\beta \hat{H}_{S}(a)}]$, $Z_{B} = \mathrm{Tr}_{B} [e^{-\beta \hat{H}_{B}^{0}}]$, and $\beta = 1/(k_{\textbf{B} } T)$ with $T$ being the temperature of the heat bath. The state of the work source and the system remains the same as the case without a bath and the bath is in a thermal state with respect to its bare Hamiltonian. For the backward process, the joint probability of measuring the transition from $\hat{\Theta} |a^{\prime},E_{m}^{a^{\prime}},E_{\nu}^{B}\rangle$ to $\hat{\Theta} |a,E_{n}^{a},E_{\mu}^{B}\rangle$ is
\begin{equation}
    P^{\text{B}}[\tilde{\Gamma}] = \left| \langle a^{\prime}, E_{m}^{a^{\prime}}, E_{\nu}^{B} | \hat{U}_{RSB} (\tau, 0) | a, E_{n}^{a}, E_{\mu}^{B} \rangle \right|^2 \langle a^{\prime},E_{m}^{a^{\prime}},E_{\nu}^{B} | \hat{\rho}_{RSB}^{\text{tr}} | a^{\prime},E_{m}^{a^{\prime}},E_{\nu}^{B} \rangle,
\end{equation}
where
\begin{equation}
    \hat{\rho}_{RSB}^{\text{tr}} = \left[ \sum_{a} p_{a}^{\text{tr}} \ket{a}\bra{a} \otimes e^{-\beta \hat{H}_{S}(a)} Z_{S}^{-1}(a) \right] \otimes e^{-\beta \hat{H}_{B}^{0}} Z_{B}^{-1},
\end{equation}
and $\hat{\Theta}\hat{\rho}_{RSB}^{\text{tr}}\hat{\Theta}^{\dagger}$ is the initial state of the backward process. We are interested in the following entropy-like quantity,
\begin{equation}
    R[\Gamma] \equiv \ln \frac{P^{\text{F}}[\Gamma]}{P^{\text{B}}[\tilde{\Gamma}]} = \beta [(E_{m}^{a^{\prime}} - E_{n}^{a}) - (E_{\mu}^{B} - E_{\nu}^{B})] + \ln\frac{Z_{S}(a^{\prime})}{Z_{S}(a)} - \ln \frac{p_{a^{\prime}}^{\text{tr}}}{p_{a}}.
\end{equation}
The change in the bare bath energy is $(E_{\nu}^{B} - E_{\mu}^{B})$. Within the weak-coupling approximation, the heat absorbed by the system is therefore $(E_{\mu}^{B}-E_{\nu}^{B})$. The change in the system energy is $(E_{m}^{a^{\prime}}-E_{n}^{a})$. The work performed by the work source on the system is then identified as
\begin{equation}
    w[\Gamma] \equiv (E_{m}^{a^{\prime}} - E_{n}^{a}) - (E_{\mu}^{B} - E_{\nu}^{B}),
\end{equation}
according to the first law of thermodynamics of the system. This identification neglects the change in the system-bath interaction energy and is therefore valid within the weak-coupling approximation. 
For a fixed value $a$ of the work-source observable, we define the equilibrium free energy associated with the conditional system Hamiltonian as $F_{S}(a) = -\beta^{-1}\ln Z_{S}(a)$. The difference between the equilibrium free energies associated with the final and initial conditional Hamiltonians is $\Delta F_{S}[\Gamma] = -\beta^{-1}\ln Z_{S}(a^{\prime}) + \beta^{-1}\ln Z_{S}(a)$. We further define $\Delta\phi[\Gamma] = -\ln p_{a^{\prime}}^{\text{tr}} + \ln p_{a}$. Choosing the distribution $p_{a^{\prime}}^{\text{tr}}$ to be the finally measured distribution of the observable $\hat{A}_{R}$ of the work source in the forward process, $\Delta\phi[\Gamma]$ denotes the distribution change of the work source during the forward process. The entropy-like $R[\Gamma]$ can therefore be written as
\begin{equation}
    R[\Gamma] = \beta w[\Gamma] - \beta \Delta F_{S}[\Gamma] + \Delta \phi [\Gamma],
\end{equation}
which has the same form as in the bath-free case (Eq.~12 in the main text). The Jarzynski-type fluctuation theorem of this quantity is obtained by calculating the following exponential average,
\begin{equation}
\begin{aligned}
    \langle e^{-R} \rangle _{\text{F}} \equiv & \sum_{R}e^{-R}P^{\text{F}}(R) = \sum_{R}e^{-R} \sum_{\Gamma}P^{\text{F}}[\Gamma]\delta(R[\Gamma] - R) \\
    =& \sum_{\Gamma}e^{-R[\Gamma]}P^{\text{F}}[\Gamma]\\
    =& \sum_{\Gamma} \frac{P^{\text{B}}[\tilde{\Gamma}]}{P^{\text{F}}[\Gamma]} P^{\text{F}}[\Gamma]\\
    =&1.
\end{aligned}
\end{equation}
The Crooks-type fluctuation theorem is obtained by combining the forward entropy-like quantity $R[\Gamma]$ and the backward entropy-like quantity $R[\tilde{\Gamma}] \equiv \ln (P^{\text{B}}[\tilde{\Gamma}]/P^{\text{F}}[\Gamma]) = -R[\Gamma]$. It is not hard to verify that
\begin{equation}
    \frac{P^{\text{F}}(R)}{P^{\text{B}}(-R)} = e^{R},
\end{equation}
where $P^{\text{F}}(R)=\sum_{\Gamma}P^{\text{F}}[\Gamma]\delta(R[\Gamma]-R)$ and $P^{\text{B}}(-R)=\sum_{\tilde{\Gamma}}P^{\text{B}}[\tilde{\Gamma}]\delta(R[\tilde{\Gamma}]+R)$.

\subsection{Exclusive work}

For the exclusive-work framework, we measure the observable $\hat{\Xi}$ of the work source, the bare Hamiltonian $\hat{H}_{S}^{0}$ of the system, and the bare Hamiltonian $\hat{H}_{B}^{0}$ of the bath at initial and final times. The joint probability of initially obtaining $\xi$ as the eigenvalue of $\hat{\Xi}$, $e_{n}$ as the eigenvalue of $\hat{H}_{S}^{0}$ and $E_{\mu}^{B}$ as the eigenvalue of $\hat{H}_{B}^{0}$, and finally obtaining $\xi^{\prime}$, $e_{m}$ and $E_{\nu}^{B}$ is
\begin{equation}
    P^{\text{F}}[\gamma] = \left| \langle \xi^{\prime}, e_{m}, E_{\nu}^{B} | \hat{U}_{RSB}(\tau, 0) | \xi, e_{n}, E_{\mu}^{B} \rangle \right|^2 \langle \xi, e_{n}, E_{\mu}^{B} | \hat{\eta}_{RSB} | \xi, e_{n}, E_{\mu}^{B} \rangle,
\end{equation}
where $\gamma$ denotes the transition from $|\xi, e_{n}, E_{\mu}^{B} \rangle$ to $| \xi^{\prime}, e_{m}, E_{\nu}^{B} \rangle$ and the initial state $\hat{\eta}_{RSB}$ reads
\begin{equation}
    \hat{\eta}_{RSB} = \sum_{\xi}p_{\xi}\ket{\xi}\bra{\xi} \otimes e^{-\beta \hat{H}_{S}^{0}} Z_{0}^{-1} \otimes e^{-\beta \hat{H}_{B}^{0}}Z_{B}^{-1},
\end{equation}
where $Z_{0} \equiv \mathrm{Tr}_{S}[e^{-\beta \hat{H}_{S}^{0}}]$. 
For the backward process, the joint probability of measuring the transition from $\hat{\Theta}|\xi^{\prime}, e_{m}, E_{\nu}^{B} \rangle$ to $\hat{\Theta}| \xi, e_{n}, E_{\mu}^{B}\rangle$ is
\begin{equation}
    P^{\text{B}}[\tilde{\gamma}] = \left| \langle \xi^{\prime}, e_{m}, E_{\nu}^{B} | \hat{U}_{RSB}(\tau, 0) | \xi, e_{n}, E_{\mu}^{B} \rangle \right|^2 \langle \xi^{\prime}, e_{m}, E_{\nu}^{B} | \hat{\eta}_{RSB}^{\text{tr}} | \xi^{\prime}, e_{m}, E_{\nu}^{B} \rangle,
\end{equation}
where
\begin{equation}
    \hat{\eta}_{RSB}^{\text{tr}} = \sum_{\xi}p_{\xi}^{\text{tr}}\ket{\xi}\bra{\xi} \otimes e^{-\beta \hat{H}_{S}^{0}} Z_{0}^{-1} \otimes e^{-\beta \hat{H}_{B}^{0}}Z_{B}^{-1}.
\end{equation}
Define the entropy-like quantity,
\begin{equation}
\begin{aligned}
    r[\gamma] \equiv& \ln \frac{P^{\text{F}}[\gamma]}{P^{\text{B}}[\tilde{\gamma}]}\\
    =& \beta[(e_{m} - e_{n}) - (E_{\mu}^{B} - E_{\nu}^{B})]-\ln\frac{p_{\xi^{\prime}}^{\text{tr}}}{p_{\xi}} \\
    =& \beta w_{\text{exc}}[\gamma] - \ln\frac{p_{\xi^{\prime}}^{\text{tr}}}{p_{\xi}},
\end{aligned}
\end{equation}
where
\begin{equation}
    w_{\text{exc}}[\gamma] \equiv (e_{m} - e_{n}) - (E_{\mu}^{B} - E_{\nu}^{B}),
\end{equation}
since $(e_{m} - e_{n})$ is the energy change of the system and $(E_{\mu}^{B} - E_{\nu}^{B})$ is identified as the heat absorbed by the system within the weak-coupling approximation. If $p_{\xi^{\prime}}^{\mathrm{tr}}$ is chosen as the final measured distribution of $\hat{\Xi}$ in the forward process, $- \ln(p_{\xi^{\prime}}^{\text{tr}} / p_{\xi})$ is the distribution change associated with the measured work-source observable. 
It is not hard to verify the Jarzynski-type fluctuation theorem
\begin{equation}
    \langle e^{-r} \rangle_{\text{F}} = 1,
\end{equation}
as well as the Crooks-type fluctuation theorem
\begin{equation}
    \frac{P^{\text{F}}(r)}{P^{\text{B}}(-r)} = e^{r}.
\end{equation}

\twocolumngrid{}


\bibliography{Refs}

\end{document}